\documentclass[manuscript,screen,nonacm]{acmart}

\AtBeginDocument{%
  }

\usepackage{braket}
\usepackage{amsmath}
\usepackage{subcaption}
\usepackage{import}
\usepackage{booktabs}
\usepackage{makecell}

\begin{document}

\title{Evaluating the Quantum Approximate Optimization Algorithms for QUBO problems Across Quantum Hardware Platforms: Performance Analysis, Challenges, and Strategies}

\author{Teemu Pihkakoski}
\email{teemu.pihkakoski@oulu.fi}
\orcid{0009-0008-4598-7271}
\affiliation{%
  \institution{Nano and Molecular Systems Research Unit, University of Oulu}
  \city{Oulu}
  \country{Finland}
}
\author{Aravind Plathanam Babu}
\email{aravind.babu@oulu.fi}
\orcid{0000-0002-7812-3284}
\affiliation{%
  \institution{Department of Instrumentation and Applied Physics,
Indian Institute of Science, Bangalore}
  \country{India.} \institution{Nano and Molecular Systems Research Unit, University of Oulu}
  \city{Oulu}  \country{Finland}
}
\author{Pauli Taipale}
\affiliation{%
  \institution{OP Financial Group}
  \city{Helsinki}
  \country{Finland}
}
\author{Petri Liimatta}
\affiliation{%
  \institution{OP Financial Group}
  \city{Helsinki}
  \country{Finland}
}
\author{Matti Silveri}
\orcid{0000-0002-6319-2789}
\affiliation{%
  \institution{Nano and Molecular Systems Research Unit, University of Oulu}
  \city{Oulu}
  \country{Finland}
}
\renewcommand{\shortauthors}{T. Pihkakoski et al.}
\begin{abstract}
Quantum computers are expected to offer  advantages in solving  optimization problems  challenging for classical computers. Quadratic Unconstrained Binary Optimization (QUBO) problems represent an important class of problems with relevance in finance and logistics. The Quantum Approximate Optimization Algorithm (QAOA) is a prominent candidate for solving QUBO problems on near-term quantum devices. In this paper, we evaluate the performance of both the standard QAOA and the adaptive derivative assembled problem tailored QAOA (ADAPT-QAOA) to solve QUBO problems of varying sizes and hardnesses for financial feature selection problems. Our main observation is that ADAPT-QAOA achieves substantially higher approximation ratios than standard QAOA for harder feature-selection problems ($\alpha=0.6$) with statistically significant improvements observed for problem sizes $n = 6$ and $14$. However, the standard QAOA remains competitive for simpler problems. Additionally, we evaluate the practical feasibility and limitations of QAOA through a hardware-aware scaling analysis based on the real-device calibration data for various hardware platforms. We estimate that standard QAOA implementation on superconducting quantum computers provides a shorter time-to-solution compared to trapped-ion devices, while trapped-ion devices  yield more favorable error rates. Our findings provide a comprehensive overview of the challenges, trade-offs, and strategies for deploying QAOA-based methods on near-term quantum hardware.
\end{abstract}




\maketitle

\section{Introduction}
Optimization problems exist in almost every field, including finance, logistics, computational biology, and materials science. For example, in finance, a feature selection problem can model selecting the most relevant variables from a large set of candidate variables to assess the risk of a loan request. A bank may consider hundreds of potential indicators for default risk, including factors such as income levels, debt-to-income ratios, and payment history, among others. However, not all variables hold equal significance, and the objective of optimization is to select a subset of features that maximizes prediction accuracy and facilitates correct decision making. Many such optimization problems can be formulated as Quadratic Unconstrained Binary Optimization (QUBO) form, represented by a matrix encoding linear terms for each variable and quadratic terms representing correlations between variables~\cite{glover_quantum_2022, mucke_feature_2023, grant_benchmarking_2021}. QUBO problems belong to the class of NP-hard problems~\cite{glover_quantum_2022}. Widely known NP-hard problems, such as the Max-Cut problem or the Travelling Salesperson Problem (TSP), can be generalized as QUBOs. There is no known polynomial-time algorithm for solving general QUBO problems deterministically. Classical numerical solvers can still find approximate solutions, but they fail or encounter convergence difficulties when the problem size increases, requiring more time and memory resources to obtain reasonable solutions. 

Various quantum algorithms  have demonstrated the potential to offer a promising alternative to classical heuristics for solving QUBO problems.
For example, Ref.~\cite{quinton_quantum_2025}  compared the performance of quantum annealing methods versus classical solvers by solving binary linear programming problems with quadratic constraints, which were converted into QUBO problems, and the quadratic constraints were handled by adding appropriate penalties to the QUBO formulation. They used  D-Wave's hybrid solver, and showed that a binary linear programming problem with approximately $600$ variables and a single quadratic constraint can be solved in a shorter time in comparison with classical methods such as CPLEX, but at the expense of some loss in solution quality. 
Similarly, gate-based approaches like the Quantum Approximate Optimization Algorithm (QAOA), a hybrid quantum-classical variational algorithm~\cite{farhi_quantum_2014} is a promising alternative. This algorithm uses parameterized quantum circuits to explore the solution space and to find high-quality approximate solutions. Many previous works have implemented methods to study the performance of QAOA in solving optimization problems on quantum simulators or real quantum computers~\cite{fitzek_applying_2024, buonaiuto_best_2023, larkin_evaluation_2022}. However, research into finding optimization problems in which QAOA could outperform classical solvers is still ongoing. For example, QAOA has been shown to scale better than branch-and-bound solvers in terms of runtime when solving Low Autocorrelation Binary Sequences (LABS) problems~\cite{shaydulin_evidence_2024}. Many variants of the QAOA have been introduced with the aim to improve the performance and scalability of the algorithm~\cite{blekos_review_2024}. Adaptive QAOA (ADAPT-QAOA) is one of these variants, and it enhances the efficiency of finding the solution by dynamically selecting quantum gates, reducing the number of variational parameters required~\cite{zhu_adaptive_2022}. This method is inspired by the adaptive variational quantum eigensolver algorithm, ADAPT-VQE~\cite{grimsley_adaptive_2019}. The choice of gates used in the standard QAOA ansatz can significantly influence the algorithm's accuracy, computational requirements, and robustness to noise, as demonstrated in Ref.~\cite{e25081238}, where different QAOA ansatzes were investigated for solving TSP instances. In Ref.~\cite{Chen:22}, the authors studied the importance of entanglement generated by the standard QAOA and ADAPT-QAOA when solving MaxCut problems. They found that the adaptive selection of two-qubit gates in ADAPT-QAOA provides more flexibility to entangle and disentangle qubits than standard QAOA, which in turn can result in a faster converge to a solution.

We still face a considerable gap between the performance of current quantum hardware and algorithms in general. To close this gap and to improve the algorithms, we need more analysis on behavior of algorithms in various quantum use-cases and quantum computer implementations. Furthermore, from the business interest, now is an excellent time for resource estimation, as well as algorithm and hardware selection studies for potential practical use-cases. To this end, in this work, we compare the performance of the standard QAOA and ADAPT-QAOA in solving QUBO problems with a target on potential hardware implementations. Our focus is on a feature selection problem inspired in its relevance in finance applications. To compare the performance of the algorithms, we use two metrics, the approximation ratio and the time-to-solution. The approximation ratio is used to assess the accuracy of the solution that the algorithm outputs, while the time-to-solution tells about the computational resources needed to run the algorithm. We examine how the performance of these algorithms varies with the size and hardness of the problem using a quantum simulator. We also incorporate calibration data and topologies from real quantum computers to further evaluate the theoretical performance and scaling of standard QAOA implementations in practical scenarios, providing insight into their potential application for real-world optimization problems now and in the near-future. We evaluate the theoretical performance of these devices by estimating the time-to-solution and total error probability when solving the QUBO problems.

The remaining content of the paper is organized as follows. First, in Section~\ref{sec:qubo-problems}, we introduce the QUBO problems and their mathematical formulation. We also specify how to formulate feature selection problems as QUBO problems. In Section~\ref{sec:qaoa-qubo-implementation}, we discuss the implementations of standard QAOA and ADAPT-QAOA to solve feature selection problems. In Section~\ref{sec:experiments}, we present experiments and results of the performance of standard QAOA and ADAPT-QAOA when solving feature selection problems, and we also estimate the theoretical performance of real quantum computers when using standard QAOA. Finally, in Section~\ref{sec:conclusions}, we close with conclusions and outlook.

\section{QUBO and feature selection problems}\label{sec:qubo-problems}
In general, Quadratic Unconstrained Binary Optimization (QUBO) problems can be represented using an upper-triangular matrix $Q$ and a decision vector $\bar{x}$~\cite{lewis_quadratic_2017}. The elements of the vector $\bar{x}$ are binary variables and they encode the solution to the problem. A general QUBO problem is solved by minimizing the objective function
\begin{align}\label{qubo_objective_function}
    f_{\mathrm{QUBO}}(\bar{x}) = \sum_i{Q_{ii} x_i} + \sum_{i < j}{Q_{ij} x_{i}x_{j}},
\end{align}
where $Q_i$ and $Q_{ij}$ are the diagonal and off-diagonal elements of the matrix $Q$. By definition QUBO forms are unconstrained. However, if needed, constraints can be handled by encoding them as additional penalty terms within the objective function~\cite{glover_quantum_2022}.

In this work, we mainly consider feature selection problems~\cite{guyon_introduction_2003}. Feature selection problems appear in applications where the goal is to identify the most relevant variables for a predictive model while removing irrelevant or redundant ones. For example, in finance, feature selection problems are used in credit scoring applications, where the most relevant variables are selected to score the credit applicants. In feature selection problems, we have a matrix in which the diagonal elements represent the cost of selecting a feature, and the off-diagonal elements represent the cost of selecting the corresponding two features, which either reward or penalize selecting the features together. The goal of the problem is to select a subset of features from all features so that the cost function $f_{\mathrm{FS}}$ is minimized. Feature selection problems are solved by minimizing the objective function
\begin{align}\label{feature_selection_objective_function}
    f_{\mathrm{FS}}(\bar{x}) = -(1 - \alpha) \sum_i{Q_{ii} x_i} + \alpha \sum_{i < j}{Q_{ij} x_{i}x_{j}},
\end{align}
where $\alpha \in [0, 1]$ is a real-valued trade-off parameter. When $\alpha$ is closer to $0$, the problem focuses more on finding relevant individual features, which are defined by the linear terms in the problem. In this case, problems with relatively low number of features are generally easy to solve. When $\alpha$ is closer to $1$, the problem rewards more on finding solutions that minimize redundancy between variables, that is, the focus is on minimizing the sum of quadratic terms in the problem. Consequently, in general, the problem becomes more combinatorial, and thus harder to solve. In this work, we examine the simplest form of feature selection problems, that is, without additional constraints. For our experiments, we consider randomly generated $Q$ matrices for the feature selection problems. For every problem instance, we use the same $Q$ matrices throughout, varying only the trade-off parameter $\alpha$.

\section{Implementing QAOA to solve feature selection problems}\label{sec:qaoa-qubo-implementation}
To solve QUBO problems using the Quantum Approximate Optimization Algorithm (QAOA), the variables of the problem are converted from binary values to eigenvalues of the $\hat{\sigma}_z$ operator ($\pm 1$), that is, the variables of the problem are mapped to be represented by the states of the qubits. This can be done by converting the QUBO objective function into a mathematically equivalent Ising Hamiltonian using a transformation of variables
\begin{align}
    x_i \rightarrow \frac{1 - \hat{Z}_i}{2},
\end{align}
where $\hat{Z}_i$ is the Pauli-Z gate acting on a qubit $i$~\cite{lucas_ising_2014}. Performing this transformation to Eq.~\eqref{feature_selection_objective_function}, we obtain the Ising Hamiltonian $\hat{H}_C$ which can be used as the cost Hamiltonian to solve the feature selection problem using QAOA,
\begin{align}\label{ising}
    \hat{H}_C = -(1 - \alpha) \sum_i{Q_{ii} \frac{1 - \hat{Z}_i}{2}} + \alpha \sum_{i < j}{Q_{ij} \frac{1 - \hat{Z}_i}{2} \frac{1 - \hat{Z}_j}{2}}.
\end{align}
Next, we cover the steps needed to implement QAOA and ADAPT-QAOA to solve optimization problems.

\subsection{Standard QAOA}
\begin{figure}[b]
  \centering
  \resizebox{\textwidth}{!}{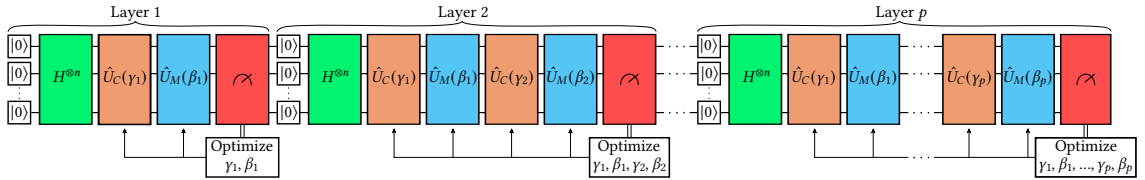}
  \caption{The standard QAOA can be divided into three parts, initialization (colored green), the application of QAOA layers (colored orange and blue), and optimization (colored red). In the initialization step, every qubit is initialized to an equal superposition state by applying Hadamard gates ($\hat{U}_H$). After the initialization, cost and mixer layers are applied to the circuit. In the optimization step, a classical optimization method is used to update the variational parameters so that the expectation value of the cost Hamiltonian with respect to the variational state is minimized. In this work, we apply the layers iteratively, optimizing all variational parameters after each addition.}\label{qaoa_circuit}
  \Description{QAOA routine.}
\end{figure}

The standard QAOA is constructed from initialization, cost and mixer layers, and optimization~\cite{farhi_quantum_2014, blekos_review_2024}. A single QAOA layer consists of a cost layer followed by a mixer layer. Formally, in a QAOA layer $k$, the cost layer $\hat{U}_C(\gamma_k)$ and the mixer layer $\hat{U}_M(\beta_k)$ are defined as
\begin{align}\label{cost_layer}
    \hat{U}_C(\gamma_k) &= e^{-i \gamma_k \hat{H}_C},  & 
    \hat{U}_M(\beta_k) &= e^{-i \beta_k \hat{H}_M},
\end{align}
where $\gamma_k$ and $\beta_k$ are variational parameters. Here $\hat{H}_M$ is the mixer Hamiltonian, which in the standard QAOA is always a rotational Pauli-X gate acting on every qubit in the circuit,
\begin{align}\label{mixer_hamiltonian}
    \hat{H}_M = \sum_{i=1}^n{\hat{X}_i},
\end{align}
where $n$ is the total number of qubits that is equal to the total number of features in the feature selection problem. Conventionally, the standard QAOA is performed by fixing the number of layers and optimizing all of the variational parameters together. We construct the standard QAOA in an iterative way, which means that we append a QAOA layer one at a time and optimizing all variational parameters after each addition, similar to what is done in ADAPT-QAOA~\cite{zhu_adaptive_2022}. This approach allows us to compare the performance of standard QAOA and ADAPT-QAOA layer by layer. To implement the first layer of the standard QAOA, we start by initializing every qubit in the circuit to an equal superposition by applying Hadamard gates ($\hat{U}_H$) as shown in Fig.~\ref{qaoa_circuit}:
\begin{align}\label{qaoa_init}
    \ket{\psi_0} = \left(\hat{U}_H \ket{0}\right)^{\otimes n}= \left( \frac{\ket{0} + \ket{1}}{\sqrt{2}} \right)^{\otimes n }.
\end{align}

After this, we apply the first QAOA layer, and the state of the system reads
\begin{align}\label{qaoa_first_layer}
    \ket{\psi_1} = e^{-i \beta_1 \hat{H}_M} e^{-i \gamma_1 \hat{H}_C} \ket{\psi_0},
\end{align}
where $\gamma_1$ is initially a random value in the range $[0, 2\pi]$ and $\beta_1$ is initially a random value in the range $[0, \pi]$. The variational parameters are updated using a classical optimization algorithm with the goal of minimizing the expectation value of the cost Hamiltonian $\hat{H}_C$ with respect to the state of the circuit. The expectation value of the cost Hamiltonian with respect to the variational state $\ket{\psi_k}$ in layer $k$ is defined as
\begin{align}\label{exp_value}
    C_k = \left\langle \psi_k \left\lvert \hat{H}_C \right\lvert \psi_k \right\rangle.
\end{align}
The optimized variational parameters of layer $k$ serve as the initial values for the variational parameters for the next layer $k+1$ with the addition of parameters $\gamma_{k+1}$ and $\beta_{k+1}$ initialized to random values in the ranges $[0, 2\pi]$ and $[0, \pi]$. The algorithm ends when all the variational parameters for the last layer $p$ have been optimized. Physically, the principle idea of QAOA is to generate a time evolution $\hat U_{\rm C}(\gamma_p)\hat U_{\rm M}(\beta_p)\ldots\hat U_{\rm C}(\gamma_1)\hat U_{\rm M}(\beta_1)$ such that it steers the ground state of the mixer Hamiltonian $\hat H_{\rm M}$ to the ground state of the cost Hamiltonian $\hat H_{\rm C}$.

The solution quality of QAOA is evaluated using the approximation ratio $r_k$ which is given by
\begin{align}
    r_k = \frac{C_k}{C_{\rm exact}},
\end{align}
where $C_k$ is the expectation value obtained at the end of layer $k$, and $C_{\rm exact}$ is the exact solution value, obtained in this work using a classical solver software Gurobi OptiMods~\cite{gurobi}. For the problem sizes that we consider in this work, the classical approach solves the problems exactly and  sufficiently fast.

\subsection{ADAPT-QAOA}
The adaptive derivative assembled problem tailored QAOA (ADAPT-QAOA) is a variant of the standard QAOA, in which the mixer Hamiltonian is adaptively chosen for each QAOA layer~\cite{zhu_adaptive_2022, blekos_review_2024}. ADAPT-QAOA is motivated by the adaptive variational quantum eigensolver algorithm, ADAPT-VQE~\cite{grimsley_adaptive_2019}. This algorithm has been shown to converge to better solutions in MaxCut problems with regular graphs than standard QAOA in the same number of layers~\cite{zhu_adaptive_2022}. The mixer for each QAOA layer is selected from a mixer pool which consists of
\begin{itemize}
  \item The standard QAOA X-mixer along with a variant that uses Y-gates, which are applied to all qubits in the circuit, $P_{\rm XY} = \left\{\sum_i^n \hat{X_i}\right\} \cup \left\{\sum_i^n \hat{Y_i}\right\}$,
  \item Single-qubit mixers, $P_{\rm single} = \bigcup_i\left\{\hat{X}_i, \hat{Y}_i\right\}$,
  \item Two-qubit mixers, $P_{\rm two} = \bigcup_{i \neq j} \left\{B_i C_j \mid B, C \in \left\{\hat{X}, \hat{Y}, \hat{Z}\right\}\right\}$.
\end{itemize}
The complete mixer pool for ADAPT-QAOA is $P_{\rm pool} = P_{\rm two} \cup P_{\rm single} \cup P_{\rm XY}$. In $P_{\rm XY}$, there are two mixers, independent of the number of qubits $n$. In $P_{\rm single}$, there are $2n$ mixers, which is straightforward to see. For the two-qubit mixers, there are $3$ choices for the operator on qubit $i$ and $3$ choices for the operator on qubit $j$, which amounts to $9$ distinct two-qubit operators on the pair $\{i, j\}$. The number of qubit pairs is ${n \choose 2} = \frac{n(n-1)}{2}$, which amounts to $9 \frac{n(n-1)}{2} = 9 \frac{n^2-n}{2}$ mixers in $P_{\rm two}$. Overall, the number of mixers in the mixer pool is $2 + 2n + 9 \frac{n^2-n}{2}$ scaling as $\mathcal{O}(n^2)$.

The mixer is selected by using a maximum gradient criterion. For QAOA layer $k$, the energy gradient for the $l$th mixer $\hat{A}_l$ in the mixer pool is given by
\begin{align}\label{grad_criterion}
    g_l = \left\lvert -i \left\langle \psi_{(k-1)} \left\lvert e^{i \hat{H}_C \gamma_0} \left[ \hat{H}_C, \hat{A}_l \right] e^{-i \hat{H}_C \gamma_0} \right\lvert \psi_{(k-1)} \right\rangle \right\rvert ,
\end{align}
where $\ket{\psi_{(k-1)}}$ is the state of the system at the end of layer $k-1$, and the parameter $\gamma_0$ is set to a small value close to zero, such as $\gamma_0 = 0.01$. The parameter $\gamma_0$ is a predefined value for the cost Hamiltonian parameter in the new layer $k$, used to evaluate the energy gradients of all candidate mixers in the mixer pool. It is important to set $\gamma_0$ close to zero to not bias the optimization, and it is the algorithmic convention to set $\gamma_0 = 0.01$~\cite{zhu_adaptive_2022}. The mixer giving the maximum gradient is selected as the mixer Hamiltonian for the layer $k$. It is believed that the mixer selection reduces excitations to higher-excited states during the algorithm improving its accuracy and speed~\cite{zhu_adaptive_2022}.

The two-qubit $\hat{Z}\hat{Z}$ mixer commutes with the cost Hamiltonian in Eq.~\eqref{ising}, and  its corresponding gradient vanishes in Eq.~\eqref{grad_criterion}. Therefore, it could be omitted from the mixer pool, decreasing the total number of mixers in the mixer pool to $2 + 2n + 8 \frac{n^2-n}{2}$, thereby further reducing the time complexity of ADAPT-QAOA. Furthermore, the Pauli operators arising from the commutator relation in Eq.~\eqref{grad_criterion} can be partitioned into mutually commuting groups and measured simultaneously, resulting in a further reduction in the measurement overhead. 
These optimizations steps were not employed in the present work but we expect that they would further reduce the cost of the ADAPT-QAOA implementation.

\section{Experiments}\label{sec:experiments}
In all experiments, we used standard QAOA or ADAPT-QAOA up to $30$ layers. We used the ideal quantum simulator of Qiskit~\cite{qiskit2024} with $10^4$ shots per optimization iteration. For the classical optimization of the variational parameters $\gamma_k$ and $\beta_k$, we used the derivative-free Powell method from the SciPy library~\cite{2020SciPy-NMeth} and  the number of iterations per layer is limited to a maximum of $1500$. We have explored various possible optimization methods, including both gradient-based and derivative-free methods. Among these, the Powell method consistently yielded the best performance and was therefore adopted throughout this work. In ADAPT-QAOA, the energy gradient is evaluated for each candidate mixer with respect to the cost Hamiltonian $\hat{H}_C$. However, this approach does not provide the complete gradient of the objective function with respect to all variational parameters, and the resulting partial gradient information may be insufficient for gradient-based optimization. Therefore, gradient-based methods may not provide any additional reduction in computational cost, and we  employ the gradient-free Powell method throughout this work. By decreasing the number of shots per optimization iteration or the maximum number of optimization iterations per layer, the algorithms can be sped up with the cost of potentially decreasing the accuracy of the algorithms. However, this does not change the scaling of time-to-solution as a function of the problem size, which is what we are interested in. For each size of the feature selection problem, we generated data for the QUBO matrices using $10$ different seeds for a random number generator, always using the same set of seeds in all experiments.

\subsection{Standard QAOA vs ADAPT-QAOA in solving feature selection problems}
\begin{figure}[t]
\centering
\begin{subfigure}{0.445\textwidth}
\centering
    $\mathbf{\alpha=0.2}$
    \vspace{-0.4cm}
    \subcaption{}
    \includegraphics[width=\linewidth]{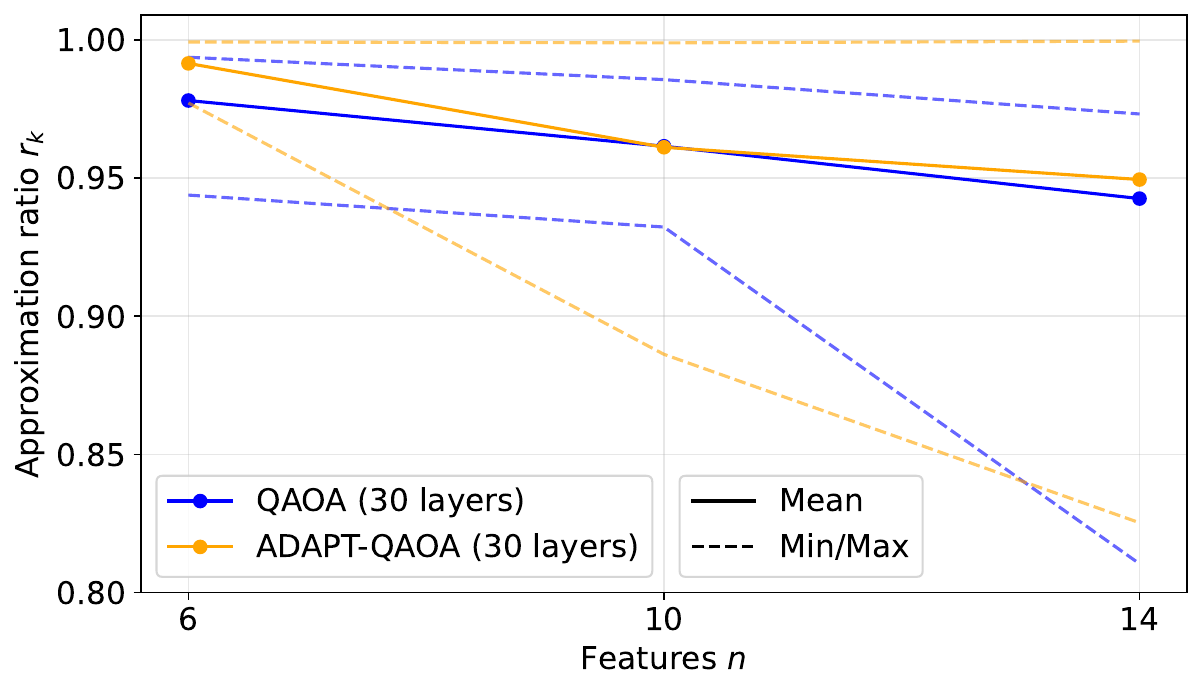}
    \label{fig:approximation_ratio_a0.2}
\end{subfigure}%
\hspace{0.5cm}
\begin{subfigure}{0.445\textwidth}
\centering
    $\mathbf{\alpha=0.6}$
    \vspace{-0.4cm}
    \subcaption{}
    \includegraphics[width=\linewidth]{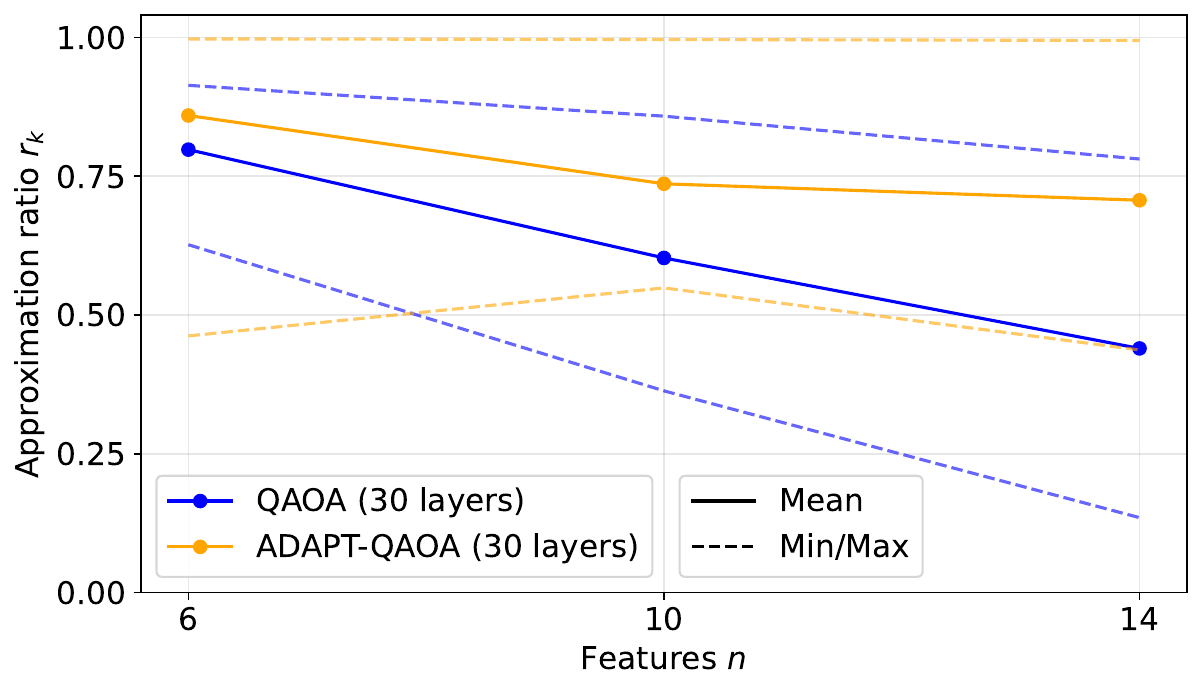}
    \label{fig:approximation_ratio_a0.6}
\end{subfigure}

\vspace{-1cm}

\begin{subfigure}{0.445\textwidth}
\centering
    \subcaption{}
    \includegraphics[width=\linewidth]{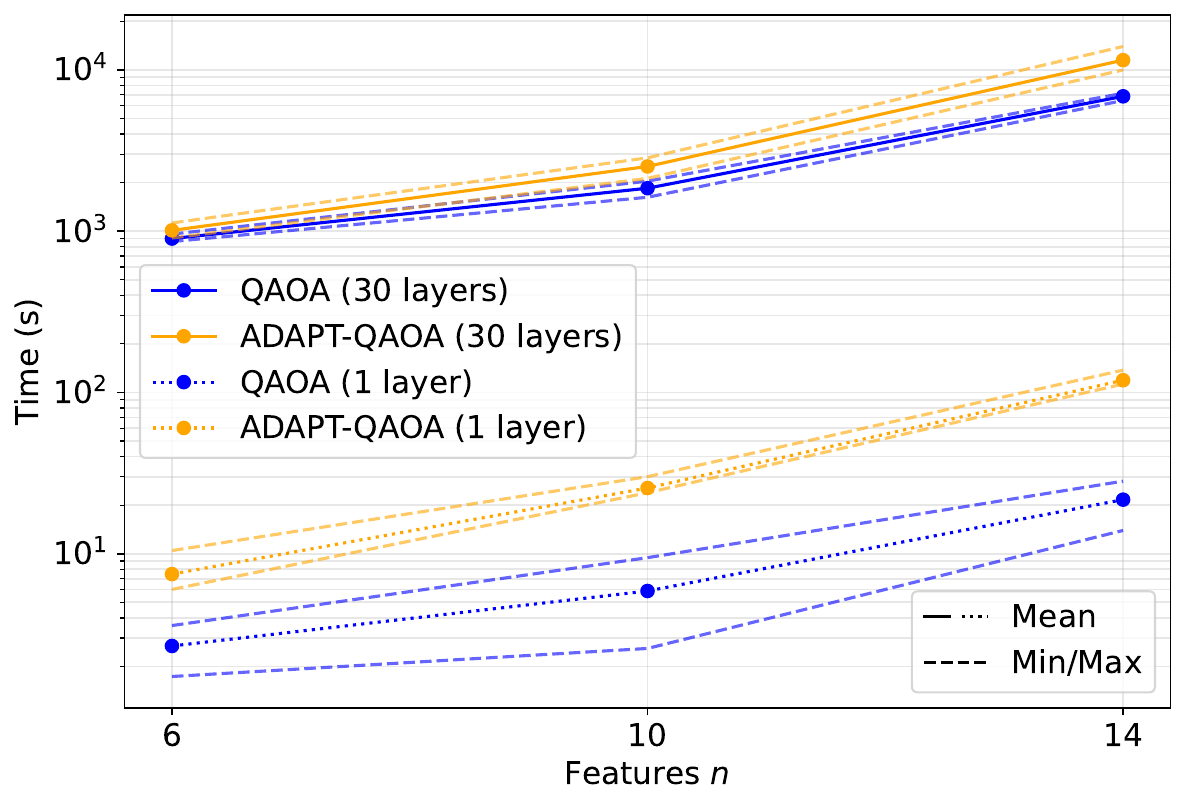}
    \label{fig:quantum_simulation_times_a0.2}
\end{subfigure}%
\hspace{0.5cm}
\begin{subfigure}{0.445\textwidth}
\centering
    \subcaption{}
    \includegraphics[width=\linewidth]{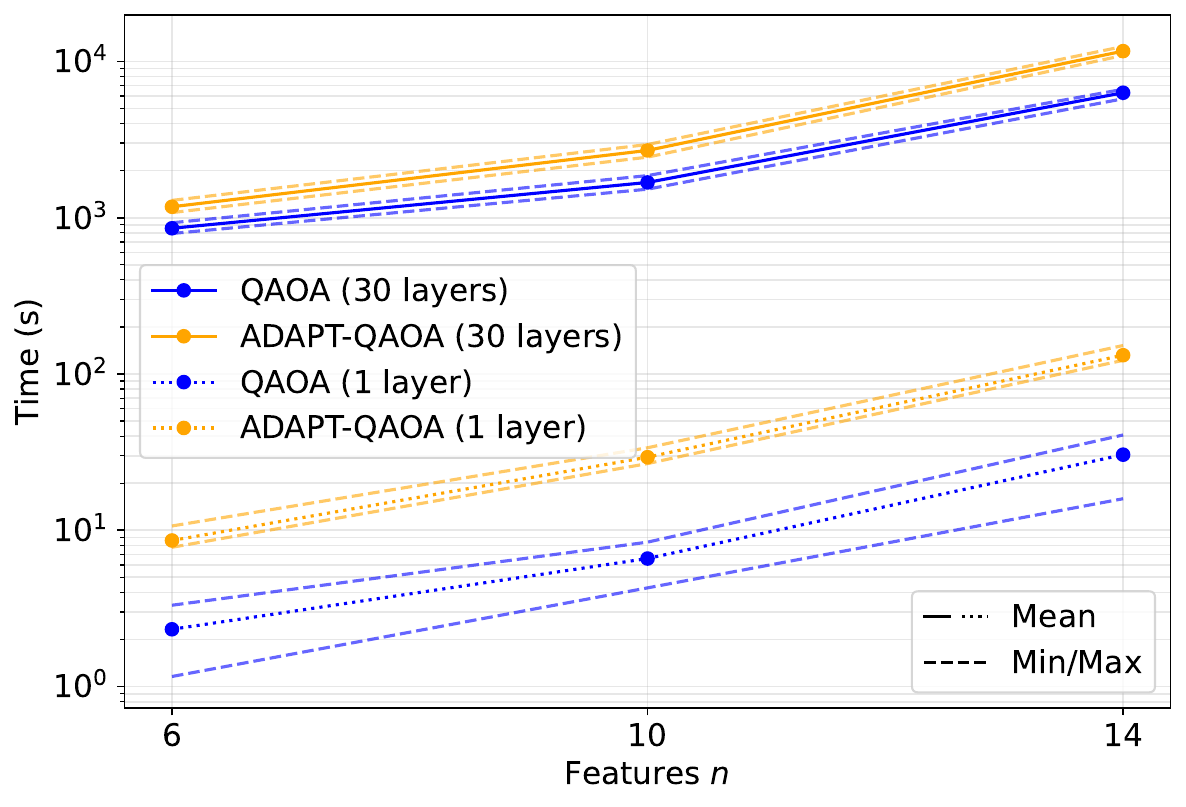}
    \label{fig:quantum_simulation_times_a0.6}
\end{subfigure}
\caption{Mean approximation ratio $r_k$ (a-b) and time-to-solution (c-d) with respect to the number of features $n$ using $30$ layers of standard QAOA and ADAPT-QAOA to solve feature selection problems with trade-off parameter set to $\alpha=0.2$ (a), (c) and $0.6$ (b), (d). Time-to-solution is given also for $1$ layer of standard QAOA and ADAPT-QAOA, for comparison. All results are averaged over $10$ problem instances obtained using $10$ different seeds to generate data for the problems. The minimum and maximum values obtained in each case are connected by dashed lines.}
\label{fig:qaoavsadaptqaoa}
\end{figure}

We used approximation ratio and time-to-solution as the metrics to compare the performance of standard QAOA and ADAPT-QAOA in solving feature selection problems. We considered problems with $6$, $10$ and $14$ features, and with $\alpha=0.2$ and $\alpha=0.6$, and the results using $30$ layers of both algorithms can be seen in Fig.~\ref{fig:qaoavsadaptqaoa}. For each problem instance, we also calculated the difference in approximation ratio between ADAPT-QAOA and standard QAOA and report the mean difference across the ten matched instances, where a positive difference means that ADAPT-QAOA achieved an better approximation ratio on average, and vice versa. The reported confidence intervals are 95\% paired $t$-intervals, calculated from the differences between the two algorithms on the same problem instances. The corresponding $p$-values are obtained using an exact two-sided paired label-swap test over the ten matched problem instances. The data for this statistical analysis is reported in Table~\ref{tab:statistical_analysis}.

\begin{table}[b]
  \caption{Statistical data for comparison of using $30$ layers of ADAPT-QAOA and standard QAOA to solve feature selection problems in terms of approximation ratio across ten matched problem instances. The table reports the mean difference in approximation ratio in approximation ratio, 95\% paired $t$-confidence interval, and exact two-sided paired label-swap test $p$-value for each problem size and $\alpha$.}
  \label{tab:statistical_analysis}
  \begin{tabular}{ccccc}
    \toprule
    $\alpha$ & $n$ & \makecell{Mean difference in\\approximation ratio} & 95\% paired CI & Two-sided $p$ \\
    \midrule
    0.2 & 6 & 0.0134 & [0.0046, 0.0222] & 0.0020 \\
    0.2 & 10 & -0.0004 & [-0.0298, 0.0290] & 0.9785 \\
    0.2 & 14 & 0.0069 & [-0.0508, 0.0646] & 0.7969 \\
    0.6 & 6 & 0.1146 & [0.0263, 0.2028] & 0.0195 \\
    0.6 & 10 & 0.1357 & [-0.0540, 0.3254] & 0.1484 \\
    0.6 & 14 & 0.1383 & [0.0269, 0.2498] & 0.0273 \\
    \bottomrule
  \end{tabular}
\end{table}

\begin{figure}[t]
\centering
\begin{subfigure}{0.445\textwidth}
\centering
    $\mathbf{\alpha=0.2}$
    \vspace{-0.4cm}
    \subcaption{}
    \includegraphics[width=\linewidth]{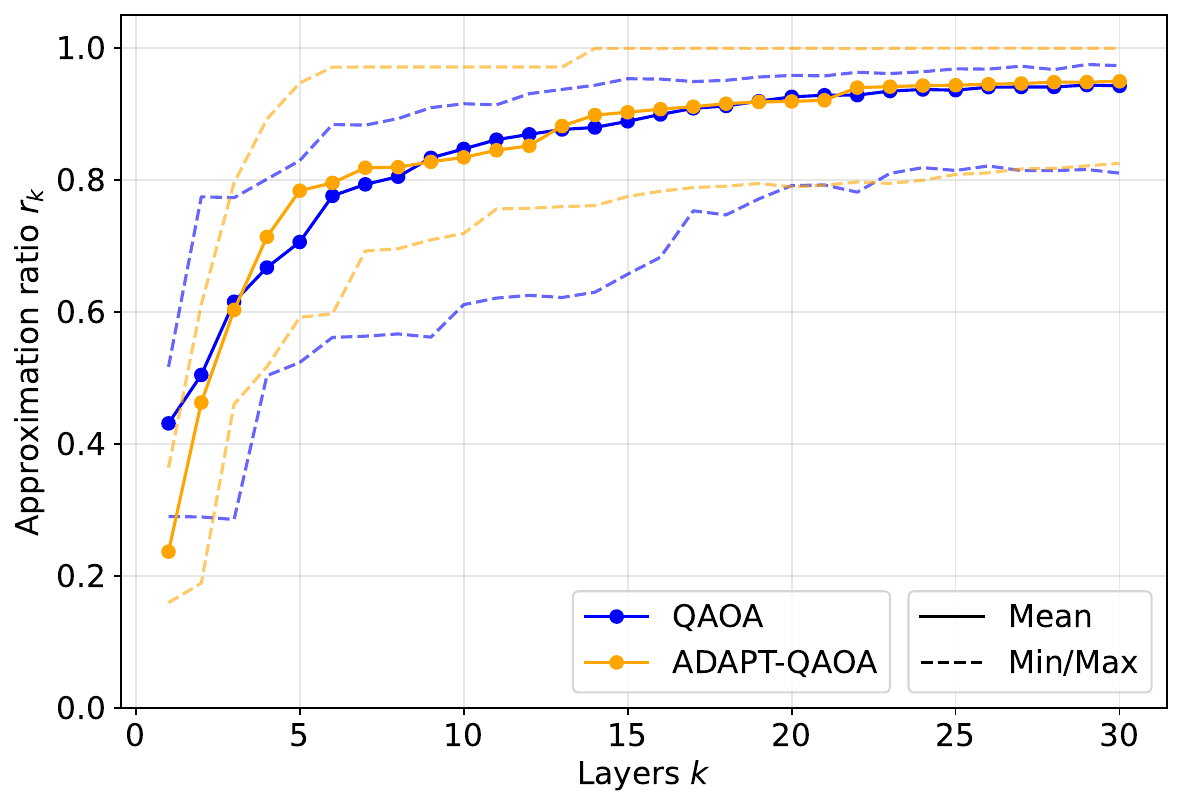}
    \label{fig:qaoavsadapteverylayer1}
\end{subfigure}%
\hspace{0.5cm}
\begin{subfigure}{0.445\textwidth}
\centering
    $\mathbf{\alpha=0.6}$
    \vspace{-0.4cm}
    \subcaption{}
    \includegraphics[width=\linewidth]{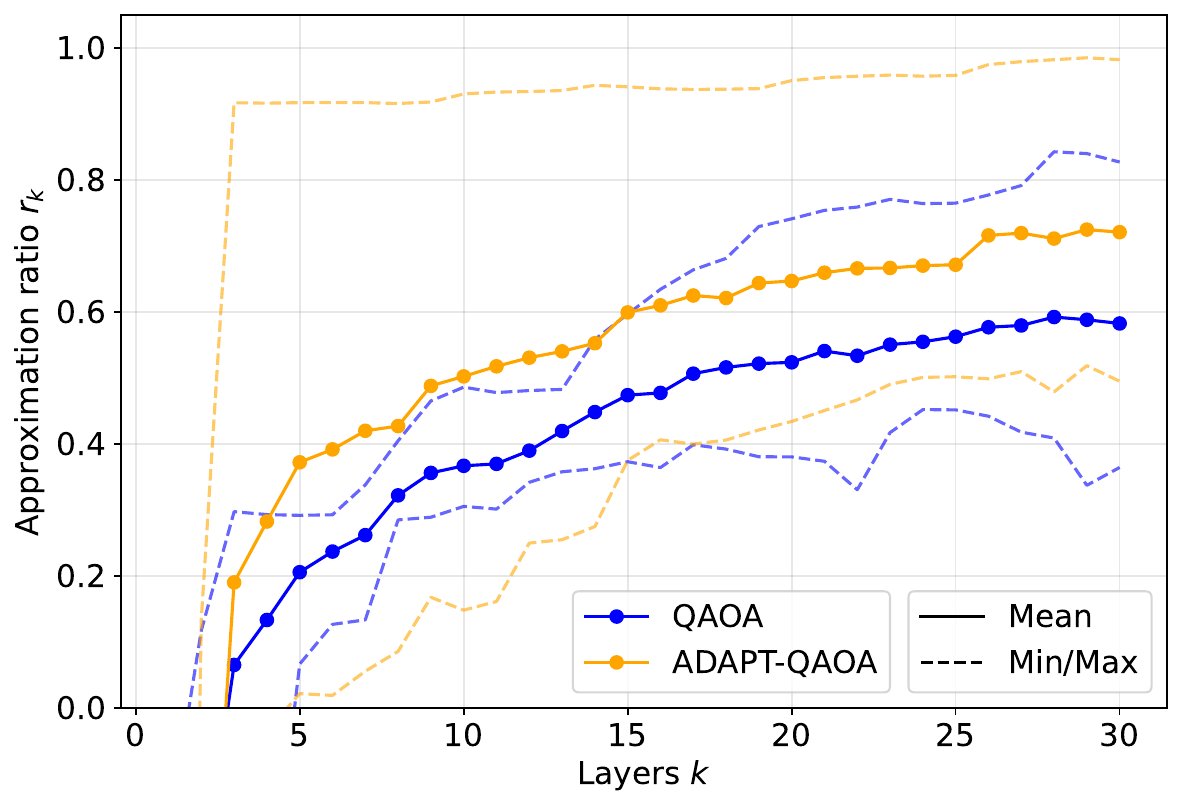}
    \label{fig:qaoavsadapteverylayer2}
\end{subfigure}

\vspace{-1cm}

\begin{subfigure}{0.445\textwidth}
\centering
    \subcaption{}
    \includegraphics[width=\linewidth]{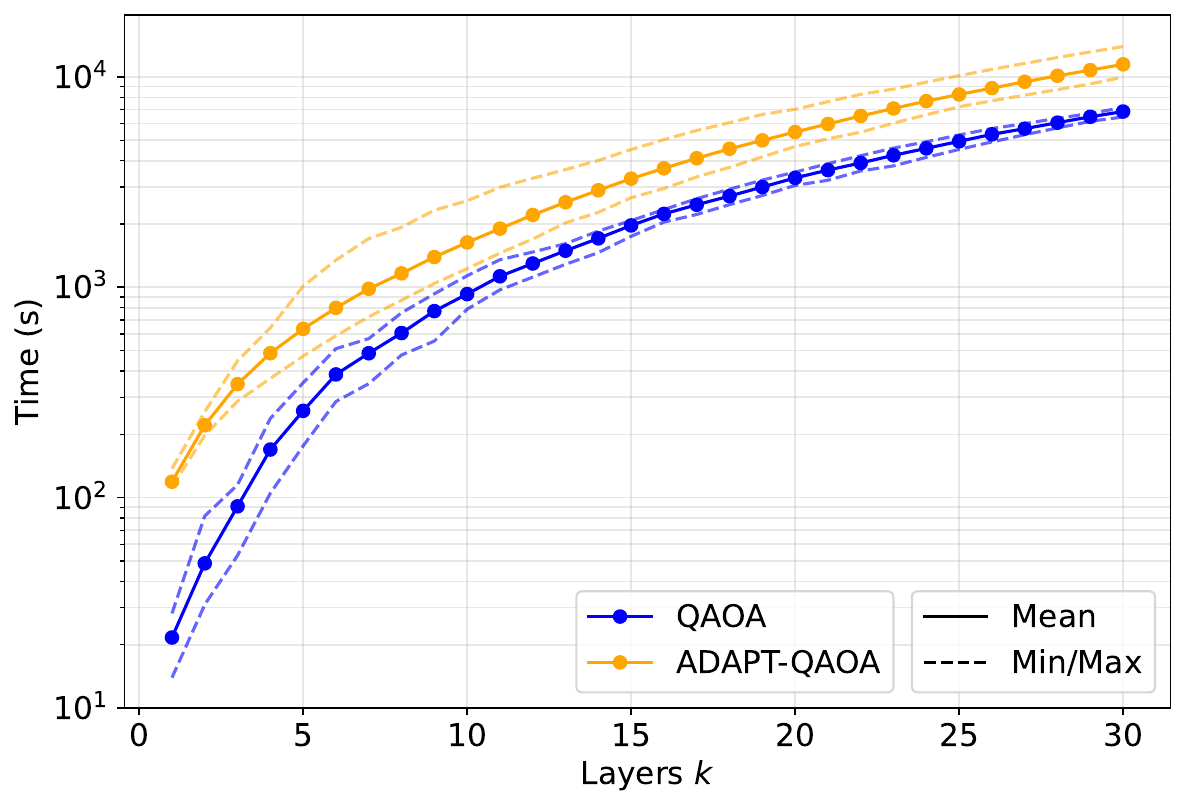}
    \label{fig:qaoavsadapteverylayer3}
\end{subfigure}%
\hspace{0.5cm}
\begin{subfigure}{0.445\textwidth}
\centering
    \subcaption{}
    \includegraphics[width=\linewidth]{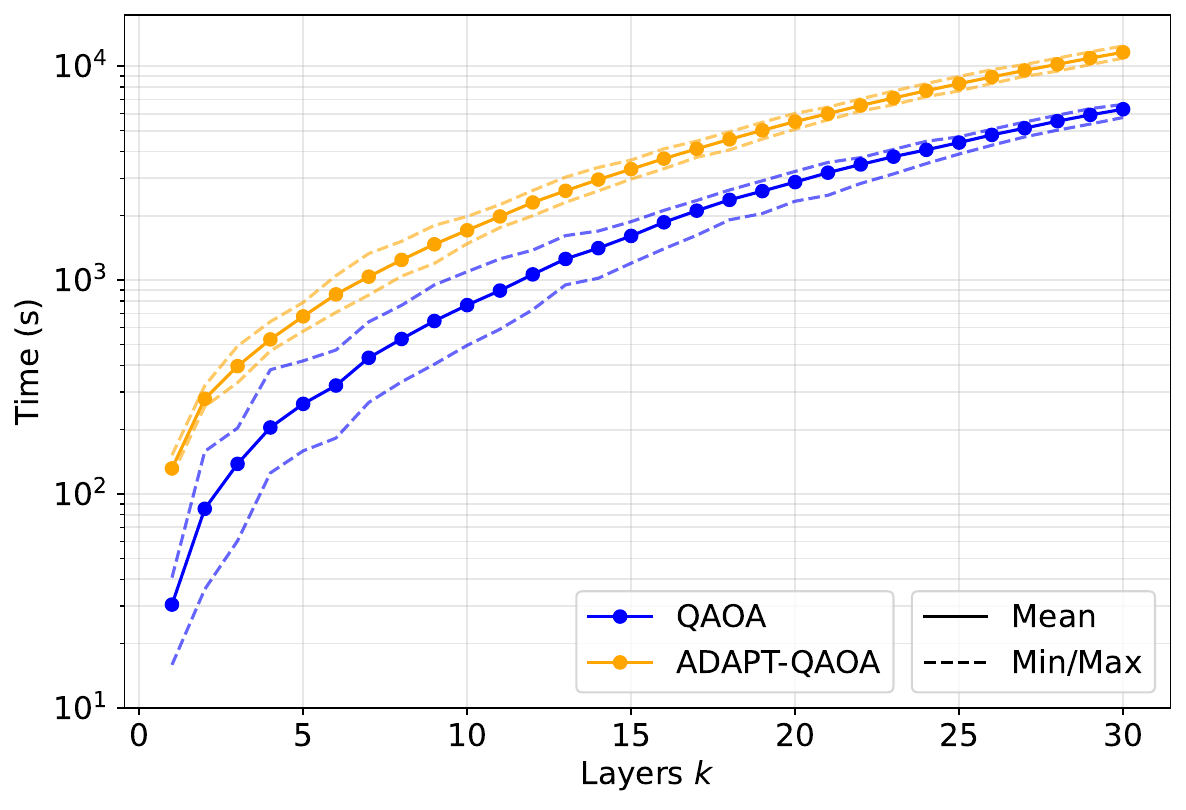}
    \label{fig:qaoavsadapteverylayer4}
\end{subfigure}
\caption{Mean approximation ratio $r_k$ (a-b) and time-to-solution (c-d) with respect to the number of layers $k$ when solving feature selection problems with $14$ features using standard QAOA and ADAPT-QAOA with trade-off parameter set to $\alpha=0.2$ (a), (c) and $0.6$ (b), (d). All results are averaged over $10$ problem instances obtained using $10$ different seeds to generate data for the problems. The minimum and maximum values obtained in each case are connected by dashed lines.}
\label{fig:qaoavsadapteverylayer}
\end{figure}

For $\alpha = 0.2$, the approximation ratios of the two algorithms are generally similar. The mean differences range from $-0.0004$ to  $0.0134$, and only the $n = 6$ case shows a statistically significant difference ($p = 0.0020$). Thus, for the easier problem setting with $\alpha = 0.2$, there is no consistent evidence of a substantial advantage for either of the algorithms in terms of approximation ratio as the problem size increases.

For $\alpha = 0.6$, the difference between the algorithms becomes more pronounced. ADAPT-QAOA achieves a higher mean approximation ratio than standard QAOA for all three problem sizes. The mean differences are $0.1146$, $0.1357$, and $0.1383$ for $n = 6$, $10$, and $14$, respectively. The differences are statistically significant for $n = 6$ ($p = 0.0195$) and $n = 14$ ($p = 0.0273$), but not for $n = 10$ ($p = 0.1484$). These results indicate that the advantage of ADAPT-QAOA in solution quality becomes more apparent for the harder problem setting with $\alpha = 0.6$. This suggests that the ability to pick a mixer more freely from a mixer pool provides ADAPT-QAOA the ability to converge to a better solution in the same amount of layers as in standard QAOA. In each of the problem sizes, there was at least one problem instance in which ADAPT-QAOA reached an approximation ratio very close to $1$, and the maximum approximation ratio of ADAPT-QAOA was always higher than that of standard QAOA.

\begin{figure}[t]
\centering
\begin{subfigure}{0.445\textwidth}
\centering
    $\mathbf{\alpha=0.2}$
    \vspace{-0.4cm}
    \subcaption{}
    \includegraphics[width=\linewidth]{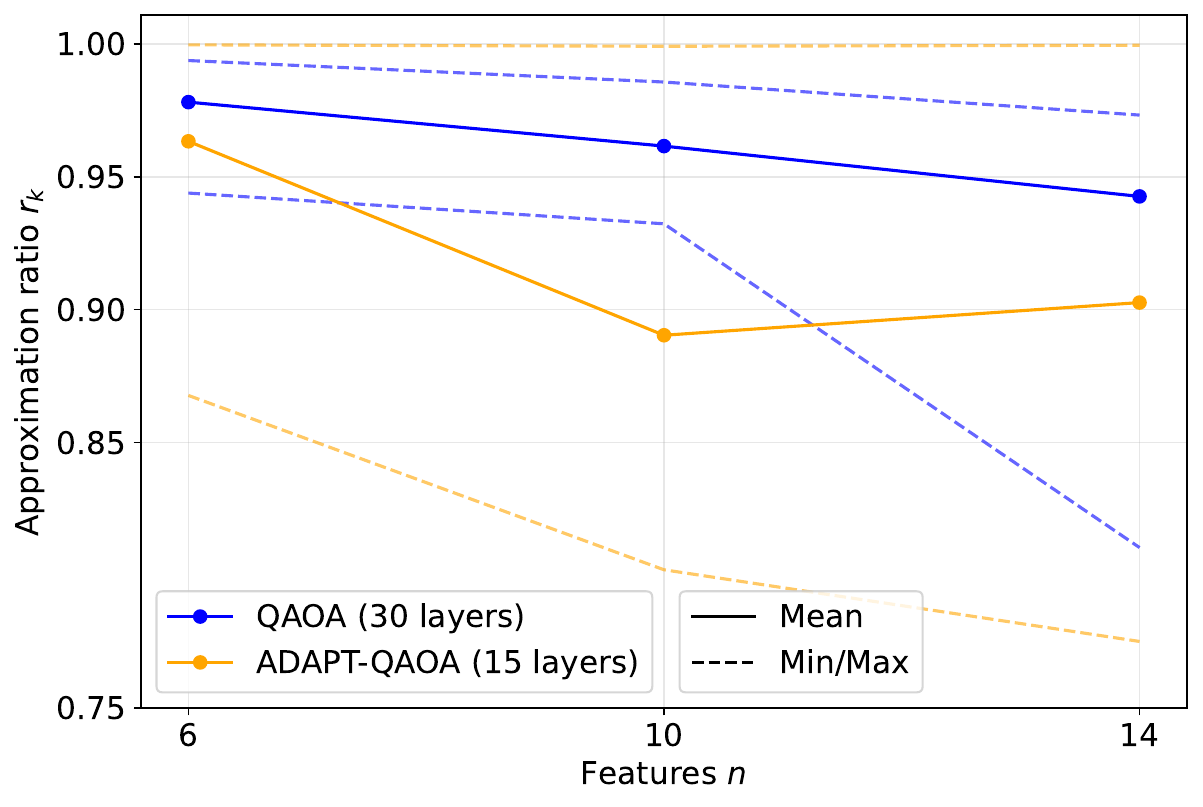}
    \label{fig:qaoavsadapt30_15_1}
\end{subfigure}%
\hspace{0.5cm}
\begin{subfigure}{0.445\textwidth}
\centering
    $\mathbf{\alpha=0.6}$
    \vspace{-0.4cm}
    \subcaption{}
    \includegraphics[width=\linewidth]{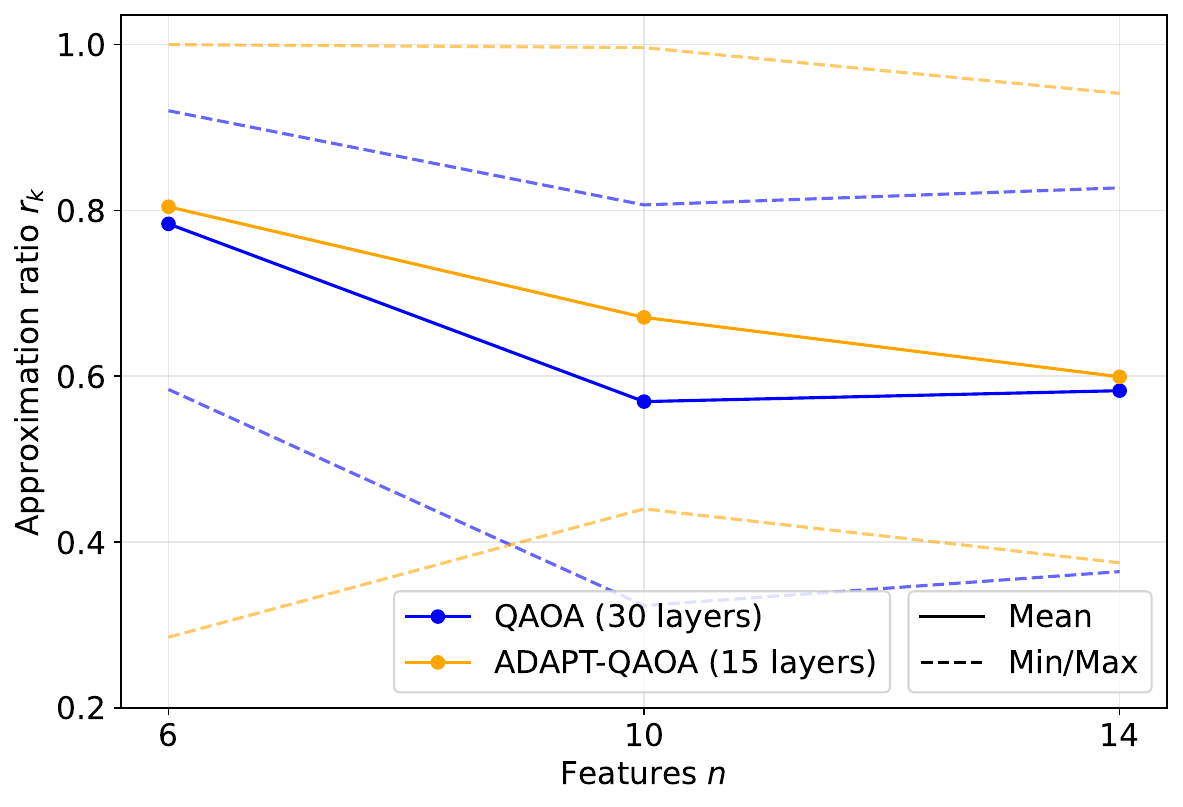}
    \label{fig:qaoavsadapt30_15_2}
\end{subfigure}

\vspace{-1cm}

\begin{subfigure}{0.445\textwidth}
\centering
    \subcaption{}
    \includegraphics[width=\linewidth]{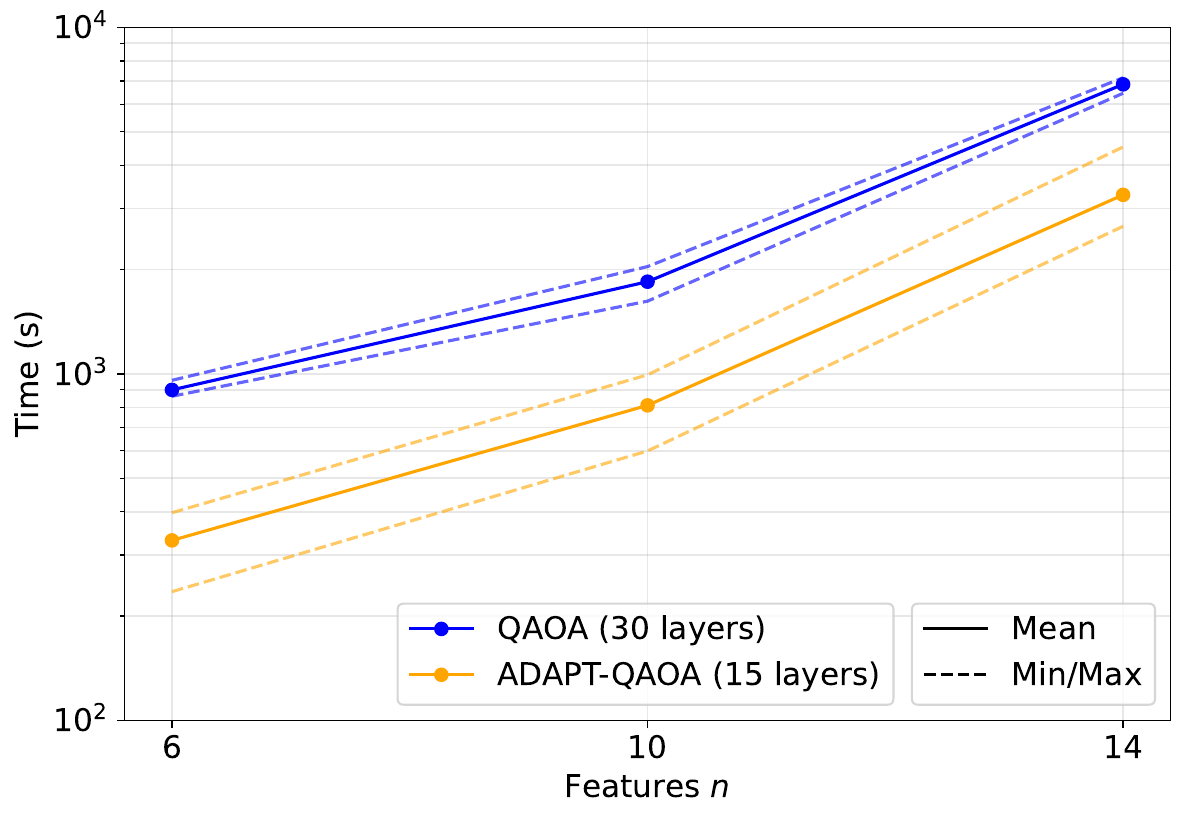}
    \label{fig:qaoavsadapt30_15_3}
\end{subfigure}%
\hspace{0.5cm}
\begin{subfigure}{0.445\textwidth}
\centering
    \subcaption{}
    \includegraphics[width=\linewidth]{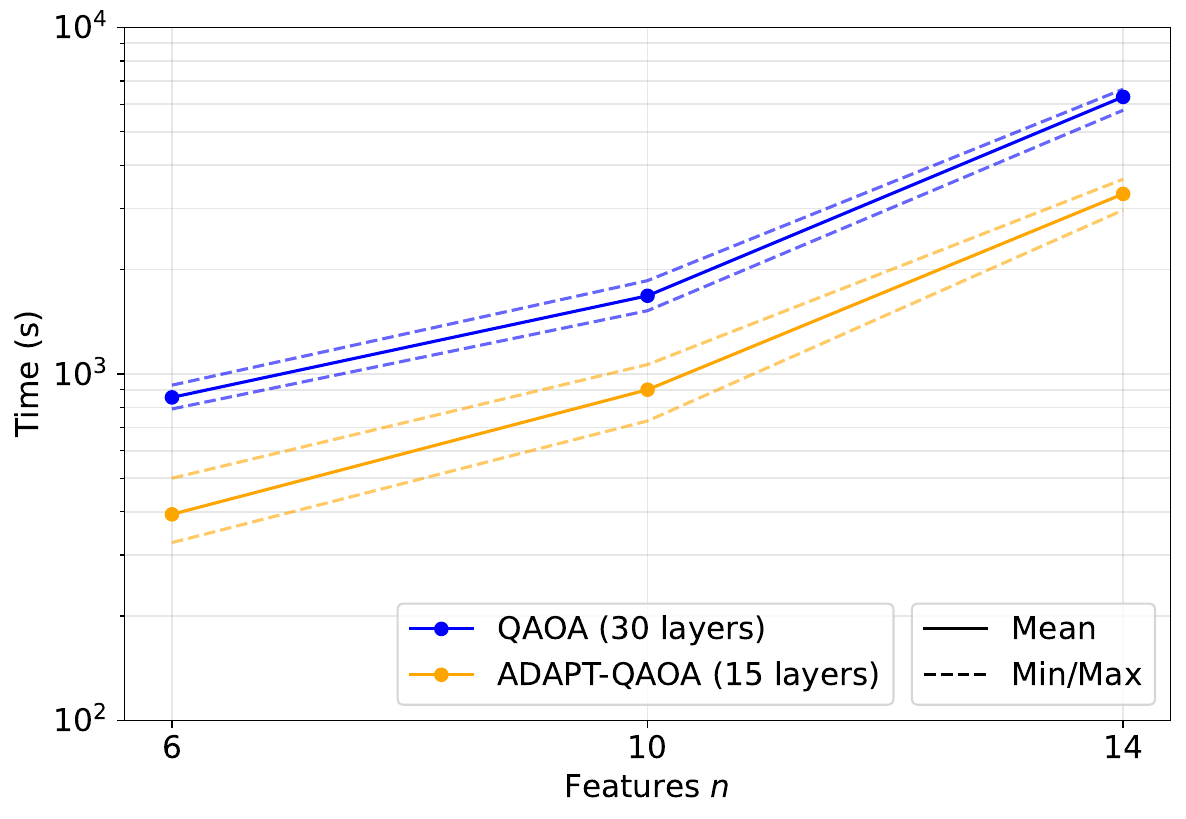}
    \label{fig:qaoavsadapt30_15_4}
\end{subfigure}
\caption{Mean approximation ratio $r_k$ (a-b) and time-to-solution (c-d) with respect to the number of features $n$ using $30$ layers of standard QAOA and $15$ layers of ADAPT-QAOA to solve feature selection problems with trade-off parameter set to $\alpha=0.2$ (a), (c) and $0.6$ (b), (d). All results are averaged over $10$ problem instances obtained using $10$ different seeds to generate data for the problems. The minimum and maximum values obtained in each case are connected by dashed lines.}
\label{fig:qaoavsadapt_30_15}
\end{figure}

\begin{table}[b]
  \caption{Statistical data for comparison of using $15$ layers of ADAPT-QAOA and $30$ layers of standard QAOA to solve feature selection problems in terms of approximation ratio across ten matched problem instances. The table reports the mean difference in approximation ratio, 95\% paired $t$-confidence interval, and exact two-sided paired label-swap test $p$-value for each problem size and $\alpha$.}
  \label{tab:statistical_analysis_15_30_layers}
  \begin{tabular}{ccccc}
    \toprule
    $\alpha$ & $n$ & \makecell{Mean difference in\\approximation ratio} & 95\% paired CI & Two-sided $p$ \\
    \midrule
    0.2 & 6 & -0.0148 & [-0.0547, 0.0252] & 0.4590 \\
    0.2 & 10 & -0.0712 & [-0.1186, -0.0239] & 0.0137 \\
    0.2 & 14 & -0.0400 & [-0.1005, 0.0206] & 0.1797 \\
    0.6 & 6 & 0.0206 & [-0.1065, 0.1478] & 0.7188 \\
    0.6 & 10 & 0.1015 & [-0.1003, 0.3034] & 0.3770 \\
    0.6 & 14 & 0.0167 & [-0.0859, 0.1194] & 0.7344 \\
    \bottomrule
  \end{tabular}
\end{table}

We next compare the two algorithms in terms of time-to-solution. With $6$ features, the time-to-solution of standard QAOA and ADAPT-QAOA are quite close to each other, but as the problem size increases, ADAPT-QAOA starts to take a longer time to solve the problem compared to standard QAOA. This arises again from the additional mixer selection step in ADAPT-QAOA, where the size of the mixer pool increases with the problem size. As a result, larger problems lead to more possible mixers to choose from in each layer, making the mixer selection process take more time. Similar kind of increase in time-to-solution is also shown in the case of $1$ layer of standard QAOA or ADAPT-QAOA as the problem size increases.

These results suggest that whenever the same amount of layers is used for both algorithms, it is preferable to use ADAPT-QAOA over standard QAOA when solution quality is prioritized over time-to-solution. This advantage becomes more pronounced as the trade-off parameter $\alpha$ increases, demonstrating that ADAPT-QAOA outperforms standard QAOA more substantially in terms of solution quality as the hardness of the problem increases.

To better highlight the performance differences between standard QAOA and ADAPT-QAOA, in Fig.~\ref{fig:qaoavsadapteverylayer} we show the approximation ratio and the time-to-solution when solving feature selection problems with $14$ features with standard QAOA or ADAPT-QAOA with layers ranging from $1$ to $30$. For example, with $\alpha=0.6$, it can be seen that, on average, $15$ layers of ADAPT-QAOA performs better than $30$ layers of standard QAOA in both approximation ratio and time-to-solution. In Fig.~\ref{fig:qaoavsadapt_30_15}, we show the results for solving feature selection problems with $6$, $10$ and $14$ features with $\alpha=0.2$ and $\alpha=0.6$ using $30$ layers of standard QAOA and $15$ layers of ADAPT-QAOA. The statistical data is reported in Table~\ref{tab:statistical_analysis_15_30_layers}. For $\alpha = 0.2$, the mean differences in approximation ratio are $-0.0148$, $-0.0712$, and $-0.0400$ for $n = 6$, $10$, and $14$, respectively. The difference is statistically significant only for $n = 10$ ($p = 0.0137$), where standard QAOA achieves a higher mean approximation ratio, and the confidence interval also indicates that standard QAOA performs better in this case. For $n = 6$ and $n = 14$, the differences are not statistically significant. For $\alpha = 0.6$, ADAPT-QAOA achieves a higher mean approximation ratio than standard QAOA for all three problem sizes, with mean differences of $0.0206$, $0.1015$, and $0.0167$ for $n = 6$, $10$, and $14$, respectively. However, none of these differences is statistically significant ($p = 0.7188$, $p = 0.3770$, and $p = 0.7344$, respectively). The corresponding confidence intervals also include zero for all three problem sizes. Thus, the observed differences in approximation ratio were not statistically significant across the ten matched instances. However, ADAPT-QAOA achieved comparable solution quality using fewer layers and a lower observed runtime.

\subsection{Classical QUBO solver}
We also solved the feature selection problems using a classical QUBO solver Gurobi OptiMods~\cite{gurobi}. This solver uses heuristic methods to provide initial solutions quickly and then uses branch-and-bound methods to explore the solution space to tighten the bounds on the optimal solution. The algorithm ends once an optimality gap is within a predefined tolerance, where the gap is defined as
\begin{align}
    \rm{gap} = \frac{|z_{\rm bound} - z_{\rm best}|}{|z_{\rm best}|},
\end{align}
where $z_{\rm best}$ is the objective value of the best solution found and $z_{\rm bound}$ is the bound on the optimal objective value. When solving the problems with this solver, we used $10^{-4}$ as the optimality gap. 

As the problem size increases, the solution space gets larger, and the classical heuristic methods are expected to become less effective at quickly finding good initial solutions. In our tests, we observed that this classical solver significantly slows down in finding the optimal solution for feature selection problems when the problem size reaches approximately $300$ variables.  This provides motivation for the future to reformulate the problem in ways that make it more challenging for classical solvers, yet still feasible to solve using the quantum algorithms.

\subsection{Comparison of different topologies and architectures}
\begin{figure}[b]
\centering
\begin{subfigure}{0.5\textwidth}
\centering
    \subcaption{}
    \includegraphics[scale=0.15]{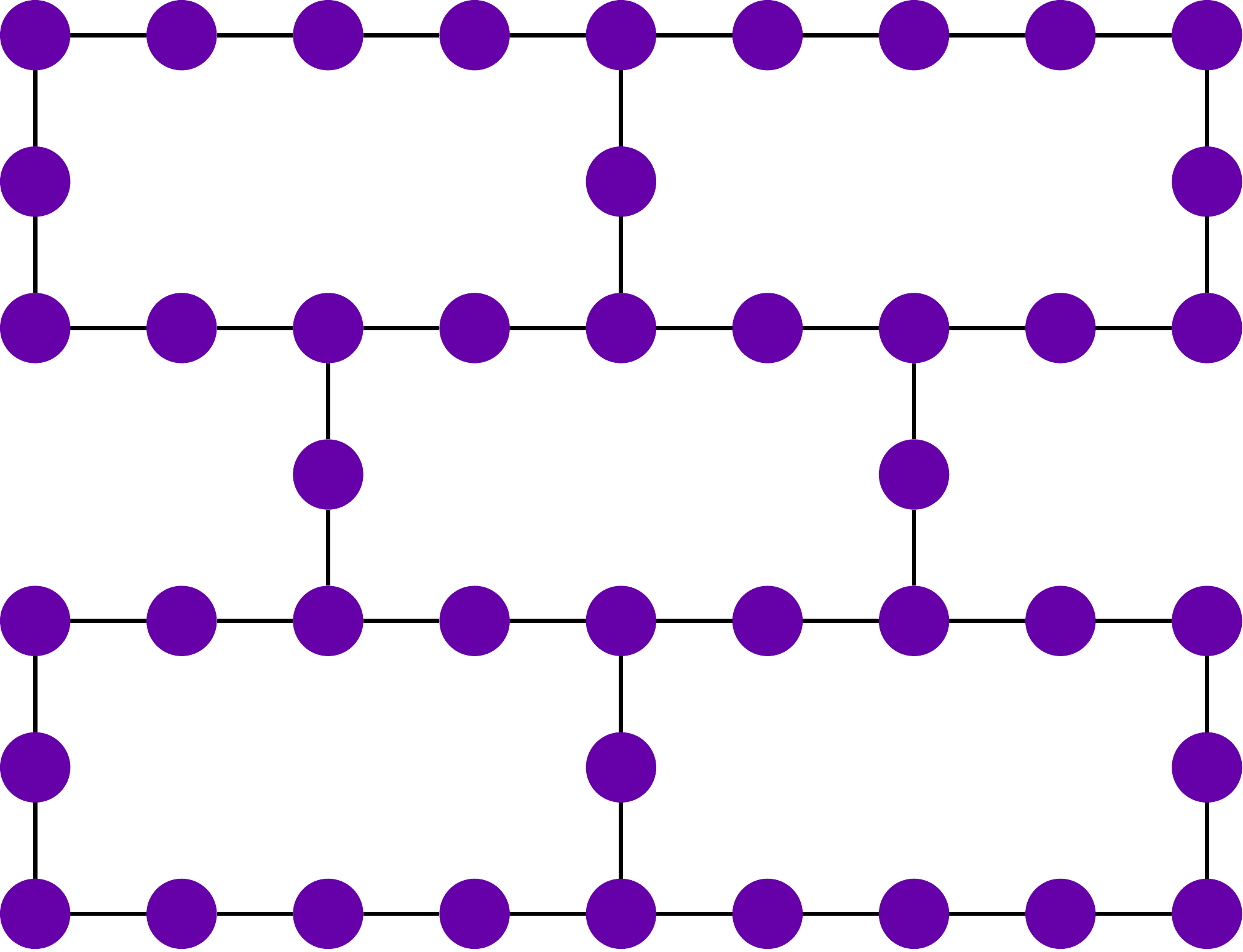}
    \label{fig:heavy_hex}
\end{subfigure}%
\begin{subfigure}{0.5\textwidth}
\centering
    \subcaption{}
    \includegraphics[scale=0.15]{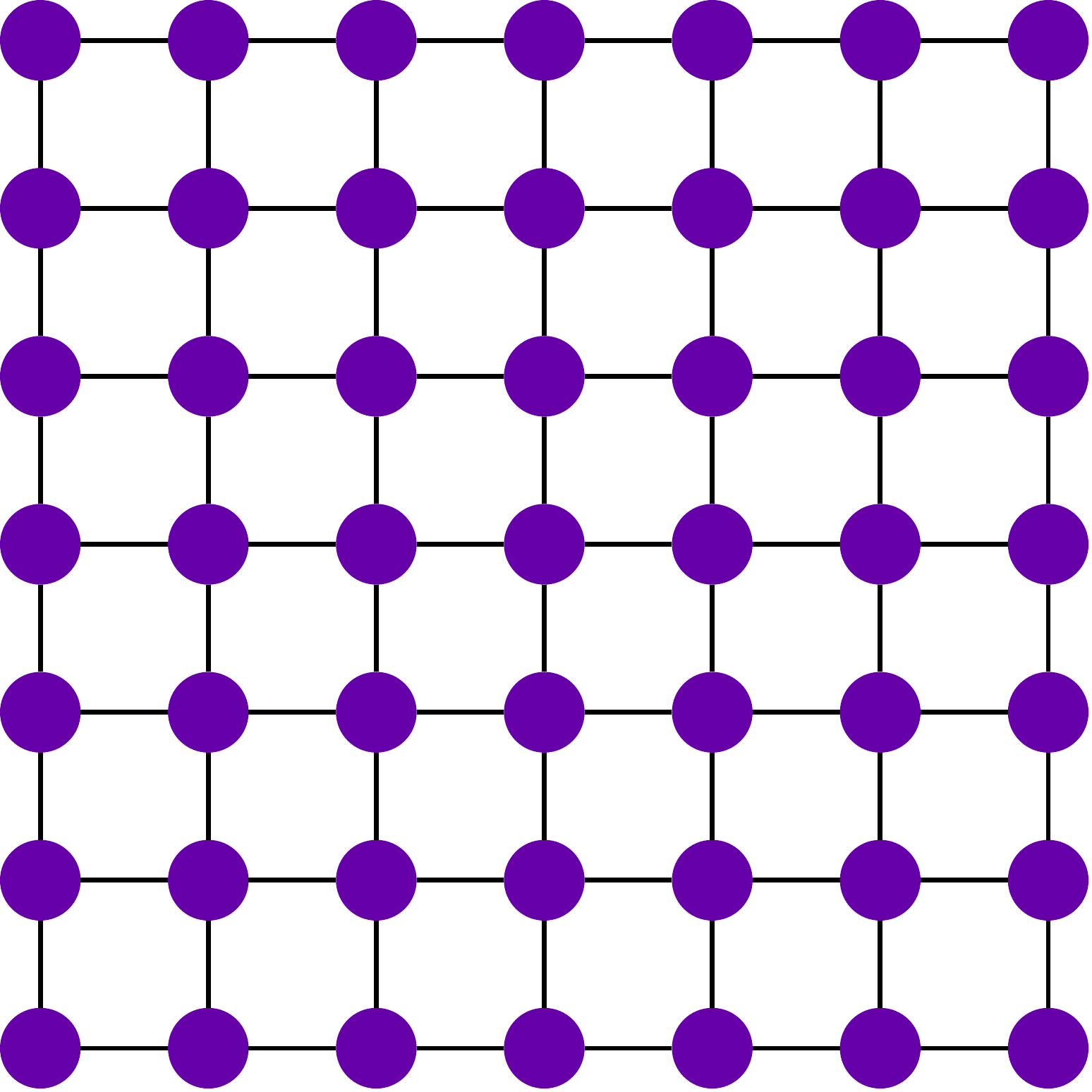}
    \label{fig:square_lattice}
\end{subfigure}
\caption{Visualization of (a) heavy-hex and (b) square lattice topologies. Purple nodes represent qubits.}
\label{fig:topologies}
\end{figure}

In order to understand how the topology and the gate durations of different quantum computers affect the performance of solving QUBO problems with QAOA, we also estimated the theoretical performance of the implementations of QAOA and ADAPT-QAOA algorithms on multiple quantum computers. The quantum computers, for which we calculated the estimates, were the Quantinuum H2 trapped-ion device~\cite{quantinuum-calibration-data} and the ibm\_brisbane superconduncting device~\cite{ibm_quantum}, which have all-to-all and heavy-hex topologies, respectively. Generally, gate operations with trapped-ion qubits are slower than with superconducting qubits, due to the additional time needed to transport qubits in trapped-ion architectures~\cite{pino_demonstration_2021}. In addition, we considered controlled connectivity scenarios based on the ibm\_brisbane device, in which the qubit connectivity is alternatively changed to a square-lattice or all-to-all topology, to investigate how qubit connectivity affects device performance. Of these topologies, the heavy-hex topology has the least connections between qubits (on average~$2.3$), the square lattice topology has more connections~($4$) than the heavy-hex topology, and lastly, in the all-to-all topology, every qubit is connected to all other qubits. Visual representations of the heavy-hex and square lattice topologies are shown in Fig.~\ref{fig:topologies}. Due to the limited amount of connections in heavy-hex and square lattice topologies, additional SWAP gates are needed to implement two-qubit gates if the qubits on which the gate operates are not connected. This generally results in a larger amount of operations needed to construct an algorithm, which in turn results in a longer time-to-solution and a higher error. Estimated time-to-solution and error probability of QAOA are chosen as the metrics to compare the performance of different quantum computers, since we want to highlight that in current devices, there is a trade-off between the time to execute the algorithm and the quality of the solution.

During the optimization part of a layer, $N$ optimization iterations are performed. In each of the optimization iterations, $S$ shots of the quantum circuit are measured. The time to execute a layer on a quantum computer can be estimated by
\begin{align}
T_{\rm layer} = (d_1 t_1 + d_2 t_2)N S,
\end{align}
where $d_1$ and $d_2$ are the number of single-qubit and two-qubit gate layers in the QAOA circuit and $t_1$ and $t_2$ are the durations of single-qubit and two-qubit gates for the quantum computer.

\begin{table}[t]
  \caption{Calibration data for ibm\_brisbane and Quantinuum H2 quantum computers. Data for ibm\_brisbane is obtained from IBM Quantum services~\cite{ibm_quantum} from a snapshot on 26 August 2025 and data for Quantinuum H2 is obtained from Refs.~\cite{pino_demonstration_2021, quantinuum-calibration-data, quantinuum-t1-t2}.}
  \label{tab:calibration_data}
  \begin{tabular}{llcc}
    \toprule
    \multicolumn{2}{l}{Value} & ibm\_brisbane & Quantinuum H2 \\
    \midrule
    Single-qubit gate duration & (\(t_1\)) & \(0.06~\mu\text{s}\) & \(63~\mu\text{s}\) \\
    Two-qubit gate duration    & (\(t_2\)) & \(0.66~\mu\text{s}\) & \(308~\mu\text{s}\) \\
    Relaxation time          & (\(T_1\)) & \(224~\mu\text{s}\) & \(4~\text{s}\) \\
    Dephasing time          & (\(T_2\)) & \(141~\mu\text{s}\) & \(60~\text{s}\) \\
    Single-qubit gate error    & (\(\epsilon_1\)) & \(3.3\times10^{-4}\) & \(3.0\times10^{-5}\) \\
    Two-qubit gate error       & (\(\epsilon_2\)) & \(1.2\times10^{-2}\) & \(1.0\times10^{-3}\) \\
    Measurement error          & (\(\epsilon_m\)) & \(3.7\times10^{-2}\) & \(1.0\times10^{-3}\) \\
    \bottomrule
  \end{tabular}
\end{table}

\begin{figure}[b]
  \centering
  \includegraphics[width=\linewidth]{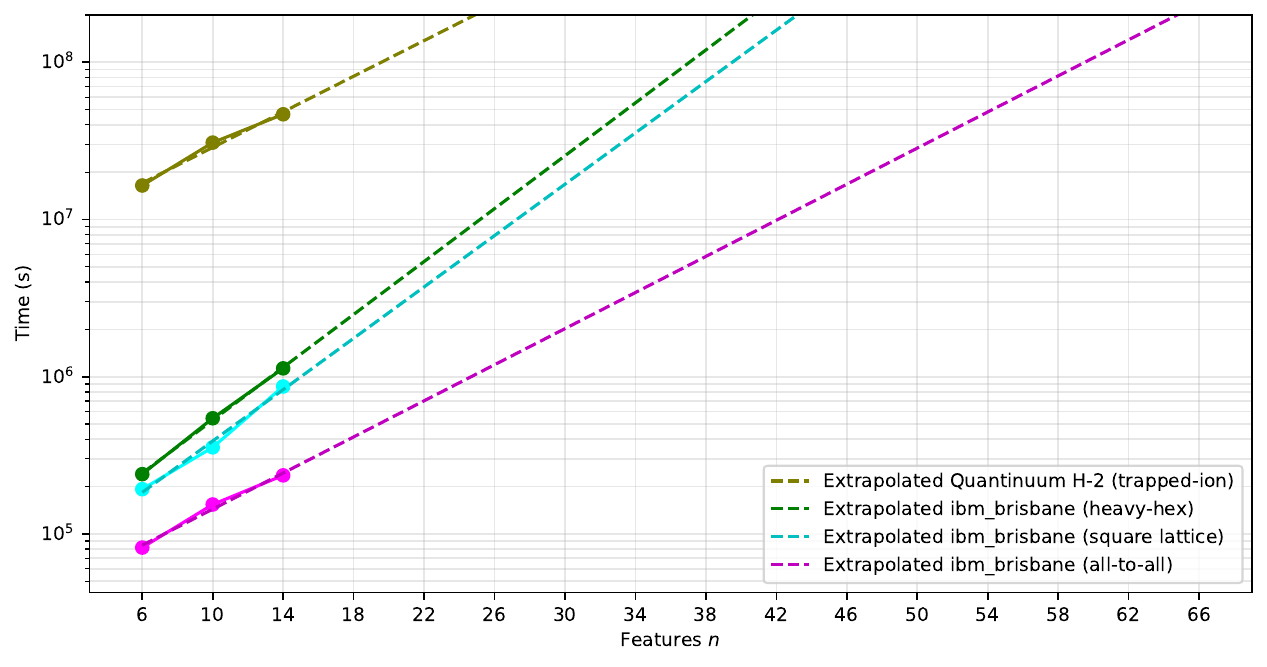}
  \caption{Mean estimated time-to-solution for feature selection problems with $\alpha = 0.2$ solved using $30$ layers of standard QAOA as a function of features $n$ for the ibm\_brisbane superconducting quantum computer and the Quantinuum H2 trapped-ion quantum computer. The ibm\_brisbane has a heavy-hex topology and the Quantinuum H2 has an all-to-all topology. In addition, we consider controlled connectivity scenarios based on the ibm\_brisbane device. The dashed lines represent extrapolations derived from the data points. All results are averaged over $10$ problem instances obtained using $10$ different seeds to generate data for the problems. 
  }
  \label{qaoa_scaling_estimates}
  \Description{Comparing different ways of solving a feature selection problem.}
\end{figure}

The number of gate layers are extracted from the quantum circuits used in the simulation of standard QAOA in solving feature selection problems by grouping gates into layers such that, within each layer, at most one gate acts on any given qubit. This ensures that all gates in the same layer can be applied simultaneously. In this estimation, the additional time due to state preparation and measurement overheads is left out, which we justify by the fact that they are approximately constant and short compared to the total time of gate durations. The gate durations $t_1$ and $t_2$ for the ibm\_brisbane quantum computer were obtained from the calibration data fetched from IBM Quantum~\cite{ibm_quantum}, and similar data for the Quantinuum H2 quantum computer was obtained from Ref.~\cite{pino_demonstration_2021}, see Table~\ref{tab:calibration_data}. For the alternative square lattice and all-to-all topologies of the ibm\_brisbane device, we use the calibration data from the real ibm\_brisbane device with a heavy-hex topology. The resulting calibration data-based error estimates are used to compare relative trends under a common error model. For the Quantinuum H2 quantum computer, the single-qubit gate duration was calculated by the sum of the intrazone shift time and the single-qubit gate time and the two-qubit gate duration was calculated by the sum of the interzone shift time and the two-qubit gate time. In the devices we consider in this work, the Z-gate can be implemented by changing the reference frame of the control electronics, thus changing the phase of all subsequent gates that act on that qubit~\cite{mckay_efficient_2017, quantinuum-z-gate}. This means that the Z-gates practically do not cost any time and do not introduce any additional error, which is taken into account in these estimations. An estimate of the total time to execute the QAOA algorithm is determined by the total time required to execute all layers.

In Fig.~\ref{qaoa_scaling_estimates}, it can be seen that with low problem sizes, the Quantinuum H2 quantum computer is estimated to take the longest amount of time to solve the problems using $30$ layers of standard QAOA. The single-qubit and two-qubit gate durations of the device  are one to two magnitudes longer than the single-qubit and two-qubit gate durations of the ibm\_brisbane quantum computer. This explains the difference in the estimated time-to-solution of the Quantinuum H2 and the ibm\_brisbane with an all-to-all alternative topology. At this stage, the Quantinuum H2 does not appear to be the most suitable quantum computer for solving these types of optimization problems in terms of time-to-solution, especially since the real device has only $56$ qubits. Compared to the ibm\_brisbane quantum computer with the heavy-hex topology, the Quantinuum H2 would need more qubits to be able to gain an advantage in time-to-solution over the ibm\_brisbane heavy-hex device. When comparing the heavy-hex and the alternative square lattice topologies on the ibm\_brisbane device, the square lattice device provides better scaling in terms of time-to-solution. This is expected since the square lattice topology has more connections between the qubits and thus needs fewer SWAP gates to implement the algorithm. Finally, of the considered quantum devices, the ibm\_brisbane quantum computer with the alternative topology of all-to-all connectivity performs the best in terms of time-to-solution.

\subsection{Error estimation}
To complement the time estimations for solving the feature selection problems using different quantum computers, we additionally estimated the errors occurring during the standard QAOA, providing a metric to evaluate and compare the solution quality achieved by the quantum computers. The total error probability~\cite{aseguinolaza_error_2024} when executing a quantum circuit can be estimated by
\begin{align}\label{error_probability_eq}
    E_{\rm tot} = 1 - \left[(1 - \epsilon_1)^{N_1} (1 - \epsilon_2)^{N_2} (1 - \epsilon_m)^{N_m} (e^{-t_c/T_1})^n (e^{-t_c/T_2})^n\right],
\end{align} 
where $\epsilon_1$, $\epsilon_2$ and $\epsilon_m$ denote the error probabilities of single-qubit gates, two-qubit gates, and measurements, respectively, $N_1$, $N_2$ and $N_M$ represent the corresponding counts of single-qubit gates, two-qubit gates, and measurements within the circuit, $t_c$ is the duration to execute the circuit, $T_1$ and $T_2$ are the relaxation and dephasing times of the device, and $n$ is the number of qubits. This error metric is constructed based on the calibration data available. It provides a sufficiently informative first-order estimate of circuit-level error by combining the reported error rates. The metric assumes that these error mechanisms contribute independently and remain approximately constant throughout circuit execution. Although this approximation is suitable for systematic circuit-level comparisons, it does not capture higher-order or context-dependent effects, including spatial and temporal crosstalk, correlated noise, calibration drift, frequency collisions, leakage, and fidelity degradation caused by simultaneous gate operations. Consequently, the metric should be interpreted as a calibration-informed estimate of circuit performance rather than a complete microscopic model of the device noise.

\begin{figure}[t]
  \centering
  \includegraphics[scale=0.6]{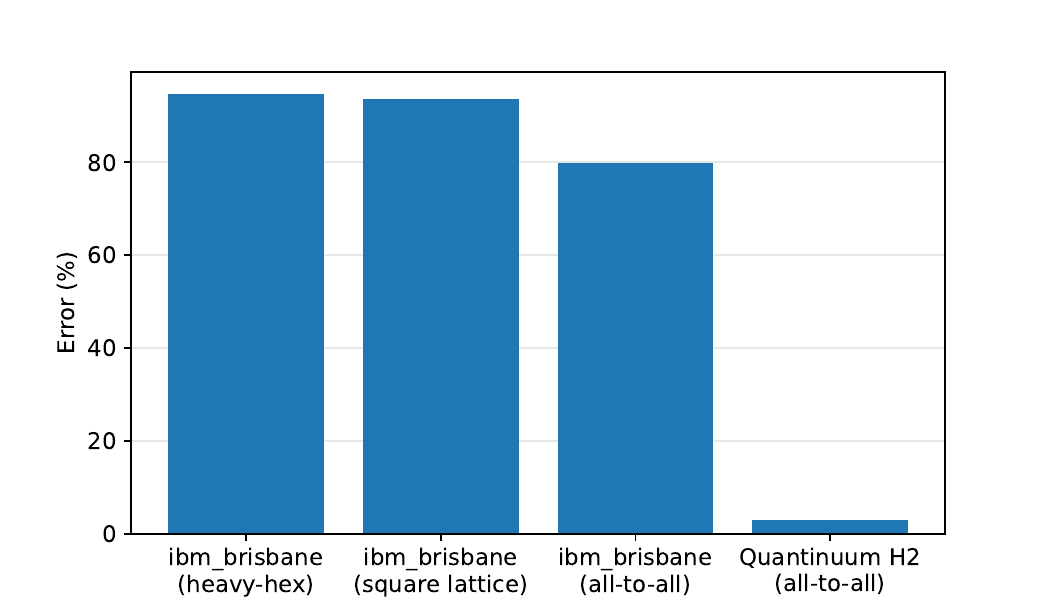}
  \caption{Estimated total error probabilities for solving a feature selection problem with $6$ features using $1$ layer of standard QAOA on the ibm\_brisbane with heavy-hex topology, hypothetical cases of ibm\_brisbane with square lattice and all-to-all topology and the Quantinuum H2 quantum computers.}
  \label{qaoa_error_estimation}
  \Description{Error estimation.}
\end{figure}

The relaxation and dephasing times and the error probabilities of the single-qubit gates, two-qubit gates, and measurements for the ibm\_brisbane quantum computer were obtained from the calibration data fetched from IBM Quantum~\cite{ibm_quantum}, and similar data for the Quantinuum H2 quantum computer was obtained from Refs.~\cite{quantinuum-calibration-data, quantinuum-t1-t2}, see Table~\ref{tab:calibration_data}. Using this information, we are using all of the available calibration data to us to estimate the error probabilities. Using Eq.~\eqref{error_probability_eq}, we estimated the total error probabilities for solving a feature selection problems with $6$ features using $1$ layer of standard QAOA. We use $1$ layer of standard QAOA to estimate the errors, since we do not consider using error mitigation methods which could have a notable impact on the errors when using, for example, $30$ layers as we did in the estimation of time-to-solution. The estimated total error probabilities are shown in Fig.~\ref{qaoa_error_estimation}. When considering the ibm\_brisbane quantum computer, the topology matters in the error estimation as topologies with fewer connections between qubits usually require more SWAP gates, and the all-to-all topology provides the lowest estimated total error probability for the ibm\_brisbane quantum computer, as expected. The Quantinuum H2 quantum computer has an all-to-all topology and its gate error rates and measurement error rates are lower and decoherence times are longer than those of the ibm\_brisbane quantum computer, and thus provides the lowest estimated total error probability for solving the feature selection problem using standard QAOA. However, as shown in Fig.~\ref{qaoa_scaling_estimates}, the time-to-solution becomes infeasible when implementing 30 layers of standard QAOA on the Quantinuum H2, although this amount of layers may be required to achieve a good approximation ratio, as shown in Fig.~\ref{fig:qaoavsadapteverylayer}. This implies that, on current devices, there is a trade-off between time-to-solution and error probability when implementing quantum algorithms.

In these error estimations, we did not consider the effect of error mitigation methods which are widely used to improve the fidelity of quantum computations when using real quantum computers. These methods include, for example, zero-noise extrapolation~\cite{li_efficient_2017} and probabilistic error cancellation~\cite{van_den_berg_probabilistic_2023}, which have their own advantages in improving the fidelity of the computations with the cost of additional classical and quantum computation time. This means that the estimated total error probabilities shown in Fig.~\ref{qaoa_error_estimation} should be considered as upper bounds for the errors. Finally, our results highlight the need for further improvements in gate error rates and measurement error rates in hardware development.

\section{Conclusions}\label{sec:conclusions}
In this work, we evaluated the performance of the standard QAOA and ADAPT-QAOA in solving QUBO problems, focusing on feature selection problems, and we estimated the performance of real quantum computers when solving the same problems using standard QAOA, taking into account the topologies, gate durations, decoherence times, and error rates of the devices.

Starting from the QUBO formulation of feature selection problems, we discussed how to encode feature selection problems into cost Hamiltonians which can be used to solve the problems using QAOA. We thoroughly explained the step-by-step implementation of standard QAOA, from initialization to optimization. We also described how to adaptively select the mixer Hamiltonian in ADAPT-QAOA using a gradient criterion.

Using an ideal simulator, we solved feature selection problem instances using standard QAOA and ADAPT-QAOA. We considered problem sizes of $6$, $10$ and $14$ features, and trade-off parameters $\alpha=0.2$ and $\alpha=0.6$, as increasing the trade-off parameter makes the problem harder. Using different seeds for a random number generator, we generated $10$ different problems for each problem size. When $\alpha = 0.2$, the two algorithms generally achieve similar approximation ratios when using $30$ layers, with no consistent evidence of an advantage for either algorithm as the problem size increases. When $\alpha = 0.6$, ADAPT-QAOA achieves higher mean approximation ratios than standard QAOA for all three problem sizes, with statistically significant improvements observed for $n = 6$ and $n = 14$, although the difference for $n = 10$ was not statistically significant. Notably, in each of the problem sizes and with both trade-off parameters, at least one instance of ADAPT-QAOA converged very close to the exact solution. With the same amount of layers, ADAPT-QAOA takes a longer time to run than the standard QAOA due to the additional mixer selection process.

We further compared $15$ layers of ADAPT-QAOA with $30$ layers of standard QAOA to investigate whether ADAPT-QAOA can achieve comparable performance with a shallower circuit. For $\alpha = 0.2$ and $0.6$, and for all problem sizes, ADAPT-QAOA achieved comparable solution quality using half the number of layers and had a lower time-to-solution. These results highlight a trade-off between solution quality and computational cost. Standard QAOA remains a competitive and computationally simpler choice for easier problems, whereas ADAPT-QAOA becomes increasingly attractive for harder problems when reducing circuit depth or improving solution quality are important considerations.

Finally, we estimated the theoretical performance of the ibm\_brisbane superconducting quantum computer and the Quantinuum H2 trapped-ion quantum computer in solving the feature selection problems using standard QAOA. We also considered alternative square lattice and all-to-all topologies on the ibm\_brisbane quantum computer. We estimated the time-to-solution and total error probabilities by extracting gate counts and circuit depth from the QAOA circuits, and by incorporating gate durations and error rates obtained from device calibration data. The gate operations of the Quantinuum H2 are relatively slow due to qubit shift operations, resulting in the longest time-to-solution in every problem size. However, the Quantinuum H2 has lower error rates and longer decoherence times than the ibm\_brisbane, and thus achieves lower total error probabilities. We also found that the topology of the device matters, as increasing connectivity between the qubits in the ibm\_brisbane results in lower time-to-solution and total error probability due to fewer SWAP operations in the quantum circuits. In the future, further improvements in qubit connectivity and gate and measurement error rates are essential to enhance the fidelity of quantum computations.

Comparing the performance of standard QAOA and ADAPT-QAOA to a classical solver Gurobi OptiMods, we found that the classical solver provides better scaling in terms of time-to-solution when solving the feature selection problems. To this end, more work is needed in finding problems that are harder to solve classically, but still tractable using quantum algorithms. An interesting direction would be to incorporate penalty terms into the QUBO objective function. Additionally, it is important to explore more efficient quantum algorithms, such as alternative variants of QAOA, and their susceptibility against various gate and qubit errors.

\begin{acks}
    We acknowledge financial support through the projects Towards Reliable Quantum Software Development: Approaches and Use Case (TORQS), grant~no.~8436/31/2022, and Securing the Quantum Software Stack (SeQuSoS), grant~no.~102/31/2024, by Business Finland; H2Future, grant no. 352788, by the Research Council of Finland and the University of Oulu; and Classi|Q⟩, project number 24304728121, jointly by the City of Oulu and the University of Oulu. A. P. B.  acknowledges the financial support from the Department of Science and Technology (DST), Government of India, through the INSPIRE Faculty Fellowship.  We are grateful for useful discussion with Matti Raasakka and all members of the TORQS project team. 
\end{acks}

\bibliographystyle{unsrturl}

\begin{thebibliography}{10}

\bibitem{glover_quantum_2022}
Fred Glover, Gary Kochenberger, Rick Hennig, and Yu~Du.
\newblock Quantum bridge analytics {I}: a tutorial on formulating and using {QUBO} models.
\newblock {\em Ann. Oper. Res.}, 314(1):141--183, July 2022.
\newblock \href {https://doi.org/10.1007/s10479-022-04634-2} {\path{doi:10.1007/s10479-022-04634-2}}.

\bibitem{mucke_feature_2023}
Sascha Mücke, Raoul Heese, Sabine Müller, Moritz Wolter, and Nico Piatkowski.
\newblock Feature selection on quantum computers.
\newblock {\em Quantum Mach. Intell.}, 5(1):11, February 2023.
\newblock \href {https://doi.org/10.1007/s42484-023-00099-z} {\path{doi:10.1007/s42484-023-00099-z}}.

\bibitem{grant_benchmarking_2021}
Erica Grant, Travis~S. Humble, and Benjamin Stump.
\newblock Benchmarking {Quantum} {Annealing} {Controls} with {Portfolio} {Optimization}.
\newblock {\em Phys. Rev. Appl.}, 15(1):014012, January 2021.
\newblock \href {https://doi.org/10.1103/PhysRevApplied.15.014012} {\path{doi:10.1103/PhysRevApplied.15.014012}}.

\bibitem{quinton_quantum_2025}
Finley~Alexander Quinton, Per Arne~Sevle Myhr, Mostafa Barani, Pedro Crespo~del Granado, and Hongyu Zhang.
\newblock Quantum annealing applications, challenges and limitations for optimisation problems compared to classical solvers.
\newblock {\em Sci. Rep.}, 15(1):12733, April 2025.
\newblock \href {https://doi.org/10.1038/s41598-025-96220-2} {\path{doi:10.1038/s41598-025-96220-2}}.

\bibitem{farhi_quantum_2014}
Edward Farhi, Jeffrey Goldstone, and Sam Gutmann.
\newblock A {Quantum} {Approximate} {Optimization} {Algorithm}, November 2014.
\newblock arXiv:1411.4028 [quant-ph].
\newblock \href {https://doi.org/10.48550/arXiv.1411.4028} {\path{doi:10.48550/arXiv.1411.4028}}.

\bibitem{fitzek_applying_2024}
David Fitzek, Toheed Ghandriz, Leo Laine, Mats Granath, and Anton~Frisk Kockum.
\newblock Applying quantum approximate optimization to the heterogeneous vehicle routing problem.
\newblock {\em Sci. Rep.}, 14(1):25415, October 2024.
\newblock \href {https://doi.org/10.1038/s41598-024-76967-w} {\path{doi:10.1038/s41598-024-76967-w}}.

\bibitem{buonaiuto_best_2023}
Giuseppe Buonaiuto, Francesco Gargiulo, Giuseppe De~Pietro, Massimo Esposito, and Marco Pota.
\newblock Best practices for portfolio optimization by quantum computing, experimented on real quantum devices.
\newblock {\em Sci. Rep.}, 13(1):19434, November 2023.
\newblock \href {https://doi.org/10.1038/s41598-023-45392-w} {\path{doi:10.1038/s41598-023-45392-w}}.

\bibitem{larkin_evaluation_2022}
Jason Larkin, Matías Jonsson, Daniel Justice, and Gian~Giacomo Guerreschi.
\newblock Evaluation of {QAOA} based on the approximation ratio of individual samples.
\newblock {\em Quantum Sci. Technol.}, 7(4):045014, August 2022.
\newblock \href {https://doi.org/10.1088/2058-9565/ac6973} {\path{doi:10.1088/2058-9565/ac6973}}.

\bibitem{shaydulin_evidence_2024}
Ruslan Shaydulin, Changhao Li, Shouvanik Chakrabarti, Matthew DeCross, Dylan Herman, Niraj Kumar, Jeffrey Larson, Danylo Lykov, Pierre Minssen, Yue Sun, Yuri Alexeev, Joan~M. Dreiling, John~P. Gaebler, Thomas~M. Gatterman, Justin~A. Gerber, Kevin Gilmore, Dan Gresh, Nathan Hewitt, Chandler~V. Horst, Shaohan Hu, Jacob Johansen, Mitchell Matheny, Tanner Mengle, Michael Mills, Steven~A. Moses, Brian Neyenhuis, Peter Siegfried, Romina Yalovetzky, and Marco Pistoia.
\newblock Evidence of scaling advantage for the quantum approximate optimization algorithm on a classically intractable problem.
\newblock {\em Sci. Adv.}, 10(22):eadm6761, May 2024.
\newblock \href {https://doi.org/10.1126/sciadv.adm6761} {\path{doi:10.1126/sciadv.adm6761}}.

\bibitem{blekos_review_2024}
Kostas Blekos, Dean Brand, Andrea Ceschini, Chiao-Hui Chou, Rui-Hao Li, Komal Pandya, and Alessandro Summer.
\newblock A review on {Quantum} {Approximate} {Optimization} {Algorithm} and its variants.
\newblock {\em Phys. Rep.}, 1068:1--66, June 2024.
\newblock \href {https://doi.org/10.1016/j.physrep.2024.03.002} {\path{doi:10.1016/j.physrep.2024.03.002}}.

\bibitem{zhu_adaptive_2022}
Linghua Zhu, Ho~Lun Tang, George~S. Barron, F.~A. Calderon-Vargas, Nicholas~J. Mayhall, Edwin Barnes, and Sophia~E. Economou.
\newblock Adaptive quantum approximate optimization algorithm for solving combinatorial problems on a quantum computer.
\newblock {\em Phys. Rev. Res.}, 4(3):033029, July 2022.
\newblock \href {https://doi.org/10.1103/PhysRevResearch.4.033029} {\path{doi:10.1103/PhysRevResearch.4.033029}}.

\bibitem{grimsley_adaptive_2019}
Harper~R. Grimsley, Sophia~E. Economou, Edwin Barnes, and Nicholas~J. Mayhall.
\newblock An adaptive variational algorithm for exact molecular simulations on a quantum computer.
\newblock {\em Nat. Commun.}, 10(1):3007, July 2019.
\newblock \href {https://doi.org/10.1038/s41467-019-10988-2} {\path{doi:10.1038/s41467-019-10988-2}}.

\bibitem{e25081238}
Wenyang Qian, Robert A.~M. Basili, Mary~Mehrnoosh Eshaghian-Wilner, Ashfaq Khokhar, Glenn Luecke, and James~P. Vary.
\newblock Comparative study of variations in quantum approximate optimization algorithms for the traveling salesman problem.
\newblock {\em Entropy}, 25(8), 2023.
\newblock \href {https://doi.org/10.3390/e25081238} {\path{doi:10.3390/e25081238}}.

\bibitem{Chen:22}
Yanzhu Chen, Linghua Zhu, Nicholas~J. Mayhall, Edwin Barnes, and Sophia~E. Economou.
\newblock How much entanglement do quantum optimization algorithms require?
\newblock In {\em Quantum 2.0 Conference and Exhibition}, page QM4A.2. Optica Publishing Group, 2022.
\newblock \href {https://doi.org/10.1364/QUANTUM.2022.QM4A.2} {\path{doi:10.1364/QUANTUM.2022.QM4A.2}}.

\bibitem{lewis_quadratic_2017}
Mark Lewis and Fred Glover.
\newblock Quadratic unconstrained binary optimization problem preprocessing: {Theory} and empirical analysis.
\newblock {\em Netw.}, 70(2):79--97, September 2017.
\newblock \href {https://doi.org/10.1002/net.21751} {\path{doi:10.1002/net.21751}}.

\bibitem{guyon_introduction_2003}
Isabelle Guyon and André Elisseeff.
\newblock An introduction to variable and feature selection.
\newblock {\em J. Mach. Learn. Res.}, 3:1157--1182, March 2003.

\bibitem{lucas_ising_2014}
Andrew Lucas.
\newblock Ising formulations of many {NP} problems.
\newblock {\em Front. Phys.}, 2, February 2014.
\newblock \href {https://doi.org/10.3389/fphy.2014.00005} {\path{doi:10.3389/fphy.2014.00005}}.

\bibitem{gurobi}
Gurobi~Optimization LLC.
\newblock {Gurobi Optimizer Reference Manual}, 2025.
\newblock URL: \url{https://www.gurobi.com/}.

\bibitem{qiskit2024}
Ali Javadi-Abhari, Matthew Treinish, Kevin Krsulich, Christopher~J. Wood, Jake Lishman, Julien Gacon, Simon Martiel, Paul~D. Nation, Lev~S. Bishop, Andrew~W. Cross, Blake~R. Johnson, and Jay~M. Gambetta.
\newblock Quantum computing with {Q}iskit, 2024.
\newblock arXiv:2405.08810 [quant-ph].
\newblock \href {https://doi.org/10.48550/arXiv.2405.08810} {\path{doi:10.48550/arXiv.2405.08810}}.

\bibitem{2020SciPy-NMeth}
Pauli Virtanen, Ralf Gommers, Travis~E. Oliphant, Matt Haberland, Tyler Reddy, David Cournapeau, Evgeni Burovski, Pearu Peterson, Warren Weckesser, Jonathan Bright, St{\'e}fan~J. {van der Walt}, Matthew Brett, Joshua Wilson, K.~Jarrod Millman, Nikolay Mayorov, Andrew R.~J. Nelson, Eric Jones, Robert Kern, Eric Larson, C~J Carey, {\.I}lhan Polat, Yu~Feng, Eric~W. Moore, Jake {VanderPlas}, Denis Laxalde, Josef Perktold, Robert Cimrman, Ian Henriksen, E.~A. Quintero, Charles~R. Harris, Anne~M. Archibald, Ant{\^o}nio~H. Ribeiro, Fabian Pedregosa, Paul {van Mulbregt}, and {SciPy 1.0 Contributors}.
\newblock {{SciPy} 1.0: Fundamental Algorithms for Scientific Computing in Python}.
\newblock {\em Nat. Methods}, 17:261--272, 2020.
\newblock \href {https://doi.org/10.1038/s41592-019-0686-2} {\path{doi:10.1038/s41592-019-0686-2}}.

\bibitem{quantinuum-calibration-data}
Quantinuum Ltd.
\newblock {System Model H2 Product Data Sheet}, 2024.
\newblock URL: \url{https://web.archive.org/web/20250318182939/https://docs.quantinuum.com/systems//data_sheets/Quantinuum%20H2%20Product%20Data%20Sheet.pdf}.

\bibitem{ibm_quantum}
{IBM Quantum}, 2021.
\newblock URL: \url{https://quantum.ibm.com/}.

\bibitem{pino_demonstration_2021}
J.~M. Pino, J.~M. Dreiling, C.~Figgatt, J.~P. Gaebler, S.~A. Moses, M.~S. Allman, C.~H. Baldwin, M.~Foss-Feig, D.~Hayes, K.~Mayer, C.~Ryan-Anderson, and B.~Neyenhuis.
\newblock Demonstration of the trapped-ion quantum {CCD} computer architecture.
\newblock {\em Nature}, 592(7853):209--213, April 2021.
\newblock \href {https://doi.org/10.1038/s41586-021-03318-4} {\path{doi:10.1038/s41586-021-03318-4}}.

\bibitem{quantinuum-t1-t2}
Quantinuum Ltd.
\newblock {Quantinuum Systems FAQs}, 2026.
\newblock URL: \url{https://web.archive.org/web/20260625064731/https://docs.quantinuum.com/systems/support/faqs.html}.

\bibitem{mckay_efficient_2017}
David~C. McKay, Christopher~J. Wood, Sarah Sheldon, Jerry~M. Chow, and Jay~M. Gambetta.
\newblock Efficient {Z} gates for quantum computing.
\newblock {\em Phys. Rev. A}, 96(2):022330, August 2017.
\newblock \href {https://doi.org/10.1103/PhysRevA.96.022330} {\path{doi:10.1103/PhysRevA.96.022330}}.

\bibitem{quantinuum-z-gate}
Quantinuum Ltd.
\newblock {System Operation}, 2024.
\newblock URL: \url{https://web.archive.org/web/20250911042853/https://docs.quantinuum.com/systems/user_guide/hardware_user_guide/system_operation.html}.

\bibitem{aseguinolaza_error_2024}
Unai Aseguinolaza, Nahual Sobrino, Gabriel Sobrino, Joaquim Jornet-Somoza, and Juan Borge.
\newblock Error estimation in current noisy quantum computers.
\newblock {\em Quantum Inf. Process.}, 23(5):181, May 2024.
\newblock \href {https://doi.org/10.1007/s11128-024-04384-z} {\path{doi:10.1007/s11128-024-04384-z}}.

\bibitem{li_efficient_2017}
Ying Li and Simon~C. Benjamin.
\newblock Efficient {Variational} {Quantum} {Simulator} {Incorporating} {Active} {Error} {Minimization}.
\newblock {\em Phys. Rev. X}, 7(2):021050, June 2017.
\newblock \href {https://doi.org/10.1103/PhysRevX.7.021050} {\path{doi:10.1103/PhysRevX.7.021050}}.

\bibitem{van_den_berg_probabilistic_2023}
Ewout van~den Berg, Zlatko~K. Minev, Abhinav Kandala, and Kristan Temme.
\newblock Probabilistic error cancellation with sparse {Pauli}–{Lindblad} models on noisy quantum processors.
\newblock {\em Nat. Phys.}, 19(8):1116--1121, August 2023.
\newblock \href {https://doi.org/10.1038/s41567-023-02042-2} {\path{doi:10.1038/s41567-023-02042-2}}.

\end{thebibliography}

\end{document}